\documentclass[12pt]{article}

\usepackage{graphicx}
\usepackage{float}
\usepackage{cite}
\usepackage{amsfonts}
\usepackage{amssymb}
\usepackage{xcolor}

\def\comp{{\rm C}\llap{\vrule height7.1pt width1pt depth-.4pt\phantom t}}
\def\gtwid{\mathrel{\raise.3ex\hbox{$>$\kern-.75em\lower1ex\hbox{$\sim$}}}}
\def\ltwid{\mathrel{\raise.3ex\hbox{$<$\kern-.75em\lower1ex\hbox{$\sim$}}}}
\def\square{\kern1pt\vbox{\hrule height 1.2pt\hbox{\vrule width 1.2pt\hskip 3pt
   \vbox{\vskip 6pt}\hskip 3pt\vrule width 0.6pt}\hrule height 0.6pt}\kern1pt}

\begin{document}

\begin{titlepage}

\begin{flushright}
UFIFT-QG-21-06 , CCTP-2021-05
\end{flushright}

\vskip 1cm

\begin{center}
{\bf Summing Inflationary Logarithms in Nonlinear Sigma Models}
\end{center}

\vskip .5cm

\begin{center}
S. P. Miao$^{1\star}$, N. C. Tsamis$^{2\dagger}$ and R. P. Woodard$^{3\ddagger}$
\end{center}

\vskip .5cm

\begin{center}
\it{$^{1}$ Department of Physics, National Cheng Kung University, \\
No. 1 University Road, Tainan City 70101, TAIWAN}
\end{center}

\begin{center}
\it{$^{2}$ Institute of Theoretical Physics \& Computational Physics, \\
Department of Physics, University of Crete, \\
GR-710 03 Heraklion, HELLAS}
\end{center}

\begin{center}
\it{$^{3}$ Department of Physics, University of Florida,\\
Gainesville, FL 32611, UNITED STATES}
\end{center}

\vspace{0.5cm}

\begin{center}
ABSTRACT
\end{center}
We consider two nonlinear sigma models on de Sitter background which
involve the same derivative interactions as quantum gravity but 
without the gauge issue. The first model contains only a single field, 
which can be reduced to a free theory by a local field redefinition; 
the second contains two fields and cannot be so reduced. Loop
corrections in both models produce large temporal and spatial 
logarithms which cause perturbation theory to break down at late times
and large distances. Many of these logarithms derive from the ``tail'' 
part of the propagator and can be summed using a variant of 
Starobinsky's stochastic formalism involving a curvature-dependent 
effective potential. The remaining logarithms derive from the 
ultraviolet and can be summed using a variant of the renormalization 
group based on a special class of curvature-dependent renormalizations. 
Explicit results are derived at 1-loop and 2-loop orders.

\begin{flushleft}
\end{flushleft}

\vskip .5cm

\begin{flushleft}
$^{\star}$ e-mail: spmiao5@mail.ncku.edu.tw \\
$^{\dagger}$ e-mail: tsamis@physics.uoc.gr \\
$^{\ddagger}$ e-mail: woodard@phys.ufl.edu
\end{flushleft}

\end{titlepage}

\section{Introduction}

In their pioneering work on perturbation theory in nontrivial geometries,
DeWitt and Brehme devoted special attention to the behavior of propagators
and Green's functions near coincidence \cite{DeWitt:1960fc}. They noted 
that while the leading singularity is a universal function of the invariant
separation, there is a less singular part which depends on the geometry and 
on the properties of the field. They called this sub-dominant singularity 
the ``tail'' term.

The tail terms of certain fields become maximally strong during the
accelerated expansion of inflation. For many purposes the background
geometry of inflation can be taken to be de Sitter,
\begin{equation}
ds^2 = a^2(\eta) \Bigl[ -d\eta^2 + d\vec{x} \!\cdot\! d\vec{x}\Bigr] =
-dt^2 + a^2 d\vec{x} \!\cdot\! d\vec{x} \quad , \quad a(\eta) = 
-\frac1{H \eta} = e^{H t} \; . \label{geometry}
\end{equation}
On $D=4$ dimensional de Sitter background the propagator of a massless,
minimally coupled scalar, in Bunch-Davies vacuum, is,
\begin{equation}
i\Delta(x;x') \Bigl\vert_{D=4} = \frac1{4 \pi^2} \Biggl\{ \frac1{a a'
\Delta x^2} - \frac{H^2}{2} \ln\Bigl[ \frac14 H^2 \Delta x^2\Bigr]
\Biggr\} \; ,  \label{propD4}
\end{equation}
where the Poincar\'e interval is,
\begin{equation}
\Delta x^2(x;x') \equiv \Bigl\Vert \vec{x} \!-\! \vec{x}' \Bigr\Vert^2 
- \Bigl( \vert \eta \!-\! \eta'\vert \!-\! i \epsilon \Bigr)^2 \; .
\label{Dx2def}
\end{equation}
The tail term of expression (\ref{propD4}) is the part involving the 
logarithm.

A curious feature of the massless, minimally coupled scalar tail is
that its coincidence limit grows with time \cite{Vilenkin:1982wt,
Linde:1982uu,Starobinsky:1982ee},
\begin{equation}
i\Delta(x;x) = \Bigl({\rm Divergent\ constant}\Bigr) + 
\frac{H^2}{4 \pi^2} \ln(a) \; . \label{IRlog1}
\end{equation}
Expression (\ref{IRlog1}) was the first example of a general sort of
secular effect encountered in loop corrections involving interactions
between nearly massless and minimally coupled scalars \cite{Onemli:2002hr,
Onemli:2004mb,Brunier:2004sb,Kahya:2006hc,vanderMeulen:2007ah}. These 
secular logarithms attracted much attention during the opening decade 
of the 21st century because they have the potential to enhance loop 
corrections to the power spectrum \cite{Boyanovsky:2005sh,Weinberg:2005vy,
Boyanovsky:2005px,Sloth:2006az,Weinberg:2006ac,Sloth:2006nu,Bilandzic:2007nb,
Seery:2007we,Seery:2007wf,Urakawa:2008rb,Riotto:2008mv,Senatore:2009cf,
Giddings:2010nc,Seery:2010kh,Kahya:2010xh}.

A fascinating aspect of the secular logarithms encountered in loop 
corrections to scalar potential models,
\begin{equation}
\mathcal{L} = -\frac12 \partial_{\mu} \Phi \partial_{\nu} \Phi g^{\mu\nu}
\sqrt{-g} - V(\Phi) \sqrt{-g} \; , \label{scalarpot}
\end{equation}
is that the steady growth of $\ln(a) = H t$ must eventually overwhelm even 
the smallest loop-counting parameter. One cannot conclude from this that 
loop corrections ever become large, just that the standard loop expansion 
breaks down. Some sort of nonperturbative resummation is required to 
determine what actually happens.

Starobinsky quite early developed a stochastic formalism which not only 
predicts the leading logarithms of scalar potential models at each order 
in perturbation theory \cite{Starobinsky:1986fx}, but also gives the late 
time form in those cases for which a static limit is approached 
\cite{Starobinsky:1994bd}. Starobinsky's formalism is based on replacing 
the full field operator $\Phi(t,\vec{x})$ with a stochastic field 
$\varphi(t,\vec{x})$ which commutes with itself $[\varphi(t,\vec{x}), 
\varphi(t',\vec{x}')] = 0$, and whose correlators are completely free of 
ultraviolet divergences. This stochastic field $\varphi(t,\vec{x})$ is 
constructed from the same free creation and annihilation operators that
appear in $\Phi(t,\vec{x})$ in such a way that the two fields produce 
the same leading logarithms at each order in perturbation theory. The 
Heisenberg field equation for $\Phi$ gives rise to a Langevin equation 
for $\varphi$ (which we express in co-moving coordinates),
\begin{equation}
\frac{\delta S[\Phi]}{\delta \Phi(x)} = \partial_{\mu} \Bigl[ \sqrt{-g} \,
g^{\mu\nu} \partial_{\nu} \Phi\Bigr] - V'(\Phi) \sqrt{-g} \longrightarrow
3 H a^3 \Bigl[ \dot{\varphi} - \dot{\varphi}_0 \Bigr] - V'(\varphi) a^3
\; . \label{exacttoLangevin}
\end{equation}
Here $\varphi_0(t,\vec{x})$ is a truncation of the Yang-Feldman free field
with the ultraviolet excised and the mode function taken to its limiting 
infrared form,
\begin{equation}
\varphi_0(t,\vec{x}) \equiv \int \!\! \frac{d^3k}{(2\pi)^3} \theta\Bigl(a H 
\!-\! k\Bigr) \frac{\theta(k \!-\! H) H}{\sqrt{2 k^3}} \Bigl\{ \alpha_{\vec{k}} 
e^{i \vec{k} \cdot\vec{x}} + \alpha^{\dagger}_{\vec{k}} e^{-i \vec{k} \cdot 
\vec{x}} \Bigr\} \; , \label{freeYF}
\end{equation}
One derives (\ref{exacttoLangevin}) by first integrating the exact field 
equation to reach the Yang-Feldman form. One then notes that reaching leading
logarithm order requires each free field to contribute an infrared logarithm, 
so there will be no change to correlators, at leading logarithm order, if the
full free field mode sum is replaced by (\ref{freeYF}). Differentiating this 
truncated Yang-Feldman equation gives Starobinsky's Langevin equation
\cite{Tsamis:2005hd}.

The problem of summing up large logarithms in flat space scattering 
amplitudes seems similar, and that has prompted particle theorists to try 
applying renormalization group methods to understanding the evolution of 
cosmological correlators \cite{Antoniadis:1991fa,Antoniadis:1992pr,
Shapiro:1999zt,Shapiro:2000dz,Bonanno:2001hi,Bonanno:2002zb,
Bentivegna:2003rr,Boyanovsky:2003ui,Reuter:2004nv,Reuter:2004nx,
Shapiro:2006qx,Sola:2007sv}. However, the problems with this approach
become obvious upon closer examination of the analogy on which it is based. 
The renormalization group of flat space describes how correlators change 
when the positions of field operators are adiabatically expanded (or 
compressed) by some constant:
\begin{equation}
{\rm Renormalization\ Group:} \qquad x^{\mu} \longrightarrow A \times 
x^{\mu} \; .
\end{equation}
What we really want to know in cosmology is how correlators change when 
{\it infinitesimal intervals} are expanded by the {\it time-dependent} 
scale factor:
\begin{equation}
{\rm Cosmological\ Evolution:} \qquad dx^{\mu} \longrightarrow a(\eta)
\times dx^{\mu} \; .
\end{equation}
It is not clear how to relate the two processes, and simple correspondences 
such as $A \longrightarrow a(\eta)$, or the renormalization scale $\mu 
\longrightarrow H$, can easily be shown to fail by direct computation 
\cite{Woodard:2008yt}. Another crucial obstacle is that the leading 
logarithms of scalar potential models arise entirely from the infrared, 
without regard to renormalization. And the fact is that, despite years of 
heroic effort by talented physicists \cite{Burgess:2009bs,Burgess:2010dd}, 
no one has yet been able to devise a version of the renormalization group 
which gives complete agreement for the leading logarithms of scalar 
potential models \cite{Burgess:2015ajz}.

Massless, minimally coupled scalars also engender large logarithms when
they interact with fermions \cite{Miao:2006pn} and with photons 
\cite{Prokopec:2002jn,Prokopec:2002uw,Prokopec:2006ue,Prokopec:2008gw}.
In both cases the other fields do not themselves generate large logarithms, 
but their dynamics modify the ways in which these logarithms manifest. 
Such modifications derive as much from the ultraviolet as from the infrared, 
so no simple truncation procedure captures the correct result. However, 
integrating out the other fields produces a scalar potential model whose 
large logarithms are correctly captured by the stochastic formalism 
\cite{Miao:2006pn,Prokopec:2007ak}.

On a general cosmological background it turns out that dynamical gravitons 
obey the same equation as the massless, minimally coupled \cite{Lifshitz:1945du},
so loop corrections from inflationary gravitons should also induce large 
logarithms. Of course the computations are much more difficult but a number 
of 1PI (one-particle-irreducible) 1-point and 2-point functions have been 
evaluated at 1-loop and 2-loop orders in pure gravity \cite{Tsamis:1996qq,
Tsamis:1996qm,Tsamis:1996qk,Tsamis:2005je} and in gravity plus various matter 
theories \cite{Miao:2005am,Miao:2012bj,Kahya:2007bc,Leonard:2013xsa,
Boran:2014xpa,Glavan:2015ura,Boran:2017fsx,Glavan:2020gal}. When the 1PI 
2-point functions are used to quantum-correct the linearized effective 
field equations one often (but not always) finds large logarithmic corrections 
to mode functions and to exchange potentials \cite{Miao:2006gj,Kahya:2007cm,
Glavan:2013jca,Wang:2014tza,Glavan:2016bvp,Glavan:2020ccz,Tan:2021lza}. These
are {\it very} challenging calculations, and it has been suggested that some 
of them may be gauge artifacts \cite{Garriga:2007zk,Tsamis:2007is,
Higuchi:2011vw,Miao:2011ng,Morrison:2013rqa,Miao:2013isa,Frob:2014fqa,
Woodard:2015kqa}. A procedure has been developed to purge gauge dependence 
from 1PI 2-point functions \cite{Miao:2017feh,Katuwal:2021ljt}, and its 
implementation is being undertaken as rapidly as the formidable computational 
challenges permit \cite{Glavan:2019msf}.

Assuming the large logarithms of inflationary gravitons are real, the question
is how they can be re-summed. The derivative couplings of gravity pose an 
obstacle to a completely stochastic explanation of these results because
derivatives preclude every free field from inducing a large logarithm, which 
was an essential part of the proof that the stochastic formalism works for 
scalar potential models \cite{Tsamis:2005hd}. Further, direct studies have 
shown that some of the logarithms cannot be explained using the stochastic 
formalism \cite{Miao:2008sp}, nor are all of the logarithms due to the tail 
part of the graviton propagator \cite{Miao:2018bol}. What we need is a simple
format in which the complications of derivative interactions can be sorted
out without intricate computations which require a year or more to complete.

Nonlinear sigma models would seem to provide a natural paradigm for 
derivative interactions. These models consist of normal scalar kinetic 
terms which are multiplied by functions of undifferentiated scalars, giving 
rise to the same sort of derivative interactions as quantum gravity but 
without the distractions of tensor indices and gauge fixing. Early work
focused on deriving a completely stochastic representation of the large
logarithms induced by these models \cite{Tsamis:2005hd}, and that approach
has been extensively pursued by Kitamoto and Kitazawa \cite{Kitamoto:2010et,
Kitamoto:2011yx,Kitamoto:2018dek}. We have thought it good to revisit this
problem after the realization that no completely stochastic approach can 
capture all the large logarithms induced inflationary gravitons 
\cite{Miao:2008sp,Miao:2018bol}. The point of this paper is to demonstrate 
that the large logarithmics of nonlinear sigma models on de Sitter can be 
explained by combining a variant of Starobinsky's stochastic formalism 
with a variant of the renormalization group.

This paper consists of six sections, of which the first is nearly done.
In section 2 we introduce the two nonlinear sigma models that will be
studied. Section 3 works out 1-loop corrections to the mode functions
and exchange potentials of the first model, as well as to 1-loop and 2-loop
expectation values of the field and its square. The same things are computed
for the second model in section 4. Section 5 collects the various large 
logarithms exposed by all this work. We then demonstrate that many of these 
large logarithms arise from stochastic effects associated with a
curvature-dependent effective potential induced by the kinetic terms. The
remaining large logarithms follow from employing the Callan-Symanzik equation
to a special class of counterterms that can be viewed as curvature-dependent
renormalizations of the bare theories. Our conclusions comprise section 6, 
particularly the lessons for quantum gravity.

\section{Two Nonlinear Sigma Models}

This section introduces the two models upon which this study is based.
The first is a single field model which gives a free theory by a local
field redefinition; the second is a model based on two fields which is
fundamentally interacting. For each model we give the bare Lagrangian
and the first two variations of the action. We also present the Feynman
rules and some important identities for the coincidence limits of the
propagator. 

\subsection{Single Field Model}

The Lagrangian of the first model we will study is,
\begin{equation}
\mathcal{L} = -\frac12 f^2(\Phi) \partial_{\mu} \Phi \partial_{\nu} \Phi
g^{\mu\nu} \sqrt{-g} \; . \label{LPhi}
\end{equation}
A nonlinear sigma model based on a single field can be reduced to free 
theories by local field redefinitions. For the case of (\ref{LPhi}) the 
free field $\Psi(x)$ obeys, 
\begin{equation}
d\Psi \equiv f(\Phi) d\Phi \qquad \Longrightarrow \qquad \mathcal{L} 
= -\frac12 \partial_{\mu} \Psi \partial_{\nu} \Psi g^{\mu\nu} \sqrt{-g} 
\; . \label{LPsi}
\end{equation}
Of course the existence of such a local field redefinition means that the
flat space $S$-matrix is unity but interactions can 
still cause interesting changes to the kinematics of free 
fields, and to the evolution of the $\Phi$ 
background. Quantifying these changes at 1-loop and 2-loop orders will 
teach us much.

We must select the function $f(\Phi)$ in order to define a specific model.
The simplest choice involves a single, dimensionful coupling constant 
$\lambda$,
\begin{equation}
f(\Phi) = 1 \!+\! \frac{\lambda}{2} \Phi \quad \Longrightarrow \quad
\Psi[\Phi] = \Phi \!+\! \frac{\lambda}{4} \Phi^2 \quad \Longleftrightarrow 
\quad \Phi[\Psi] = \frac2{\lambda} \Bigl[ \sqrt{1 \!+\! \lambda \Psi} \!-\! 
1\Bigr] \; . \label{PsiofPhi}
\end{equation}
With this choice of $f(\Phi)$ the Heisenberg field equation is,
\begin{equation}
\frac{\delta S[\Phi]}{\delta \Phi(x)} = \Bigl(1 \!+\! \frac12 \lambda 
\Phi\Bigr) \partial_{\mu} \Bigl[(1 \!+\! \frac12 \lambda \Phi \Bigr) 
\sqrt{-g} g^{\mu\nu} \partial_{\nu} \Phi\Bigr] \; . \label{varPhi1}
\end{equation}
We will also sometimes need the second variation,
\begin{eqnarray}
\lefteqn{\frac{\delta^2 S[\Phi]}{\delta \Phi(x) \delta \Phi(x')} = \lambda 
\delta^D(x \!-\! x') \partial_{\mu} \Bigl[ \Bigl(1 \!+\! \frac12 \lambda 
\Phi \Bigr) \sqrt{-g} g^{\mu\nu} \partial_{\nu} \Phi\Bigr] } \nonumber \\
& & \hspace{-0.5cm} -\frac14 \lambda^2 \delta^D(x \!-\!x') \sqrt{-g} 
g^{\mu\nu} \partial_{\mu} \Phi \partial_{\nu} \Phi + \partial_{\mu} 
\Bigl[ \Bigl(1 \!+\! \frac12 \lambda \Phi\Bigr)^2 \!\! \sqrt{-g} g^{\mu\nu} 
\partial_{\nu} \delta^D(x \!-\! x') \Bigr] . \label{varPhi2} \qquad
\end{eqnarray}

In $D$ spacetime dimensions the propagators of both the $\Phi$ and the $\Psi$ 
fields obey the same equation,
\begin{equation}
\partial^{\mu} \Bigl[a^{D-2} \partial_{\mu} i\Delta(x;x')\Bigr] \equiv
\mathcal{D} i\Delta(x;x') = i\delta^D(x \!-\! x') \; . \label{propeqn}
\end{equation}
The solution is \cite{Onemli:2002hr,Onemli:2004mb},
\begin{equation}
i\Delta(x;x') = F\Bigl( y(x;x')\Bigr) + k \ln(a a') \qquad , \qquad k \equiv
\frac{H^{D-2}}{(4\pi)^{\frac{D}2}} \frac{\Gamma(D \!-\! 1)}{\Gamma(\frac{D}2)} 
\; , \label{propagator}
\end{equation}
where the de Sitter length function is $y(x;x') \equiv a a' H^2 \Delta x^2(x;x')$
and the first derivative of the function $F(y)$ is,
\begin{eqnarray}
\lefteqn{F'(y) = -\frac{H^{D-2}}{4 (4 \pi)^{\frac{D}2}} \Biggl\{ \Gamma\Bigl(
\frac{D}2\Bigr) \Bigl( \frac{4}{y}\Bigr)^{\frac{D}2} + \Gamma\Bigl( \frac{D}2 
\!+\! 1\Bigr) \Bigl( \frac{4}{y}\Bigr)^{\frac{D}2 - 1} } \nonumber \\
& & \hspace{2cm} + \sum_{n=0}^{\infty} \Biggl[ \frac{\Gamma(n \!+\! 
\frac{D}2 \!+\! 2)}{\Gamma(n \!+\! 3)} \Bigl( \frac{y}{4}\Bigr)^{n-\frac{D}2 + 2} 
- \frac{\Gamma(n \!+\! D)}{\Gamma(n \!+\! \frac{D}2 \!+\! 1)} \Bigl( 
\frac{y}{4}\Bigr)^{n} \Biggr] \Biggr\} . \qquad \label{Fprime}
\end{eqnarray}
The coincidence limits of the propagator and its first two derivatives are,
\begin{eqnarray}
i\Delta(x;x) = k \Bigl[-\pi {\rm cot}\Bigl( \frac{D \pi}{2}\Bigr) + 2 \ln(a)\Bigr] 
& , & \partial_{\mu} i\Delta(x;x')\Bigl\vert_{x' = x} = k H a \delta^0_{~\mu} \; , 
\qquad \label{coincprop} \\
\partial_{\mu} \partial'_{\nu} i\Delta(x;x')\Bigl\vert_{x' = x} = 
-\Bigl(\frac{D\!-\!1}{D} \Bigr) k H^2 g_{\mu\nu} & , & \partial_{\mu} i\Delta(x;x) 
= 2 k H a \delta^{0}_{~\mu} \; . \qquad \label{coincddprop}
\end{eqnarray}

\subsection{Two Field Model}

The simplest truly interacting nonlinear sigma model would seem to be,
\begin{equation}
\mathcal{L} = -\frac12 \partial_{\mu} A \partial_{\nu} A g^{\mu\nu} \sqrt{-g} -
\frac12 \Bigl(1 \!+\! \frac12 \lambda A\Bigr)^2 \partial_{\mu} B \partial_{\nu} B
g^{\mu\nu} \sqrt{-g} \; . \label{LAB}
\end{equation}
The first variations of its action are,
\begin{eqnarray}
\frac{\delta S[A,B]}{\delta A(x)} & = & \partial_{\mu} \Bigl[ \sqrt{-g} g^{\mu\nu}
\partial_{\nu} A\Bigr] - \frac12 \lambda \Bigl(1 \!+\! \frac12 \lambda A\Bigr)
\partial_{\mu} B \partial_{\nu} B g^{\mu\nu} \sqrt{-g} \; , \qquad \label{Avar1} \\
\frac{\delta S[A,B]}{\delta B(x)} & = & \partial_{\mu} \Bigl[ \Bigl(1 \!+\! \frac12 
\lambda A\Bigr)^2 \sqrt{-g} g^{\mu\nu} \partial_{\nu} B\Bigr] \; . \qquad 
\label{Bvar1}
\end{eqnarray}
And the second variations work out to be,
\begin{eqnarray}
\frac{\delta^2 S[A,B]}{\delta A(x) \delta A(x')} &\!\!\! = \!\!\!& \partial_{\mu} 
\Bigl[ \sqrt{-g} g^{\mu\nu} \partial_{\nu} \delta^D(x \!-\!x') \Bigr] \nonumber \\
& & \hspace{3.5cm} - \frac14 \lambda^2 \delta^D(x \!-\! x') \partial_{\mu} B 
\partial_{\nu} B g^{\mu\nu} \sqrt{-g} \; , \qquad \label{Avar2} \\
\frac{\delta^2 S[A,B]}{\delta B(x) \delta B(x')} &\!\!\! = \!\!\!& \partial_{\mu} 
\Bigl[ \Bigl(1 \!+\! \frac12 \lambda A\Bigr)^2 \sqrt{-g} g^{\mu\nu} \partial_{\nu} 
\delta^D(x \!-\! x') \Bigr] \; . \qquad \label{Bvar2}
\end{eqnarray}

\begin{figure}[H]
\centering
\includegraphics[width=11cm]{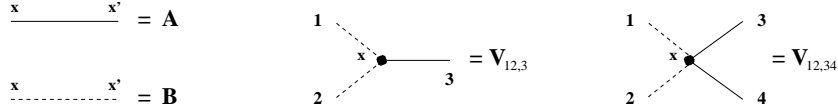}
\caption{\footnotesize The Feynman rules of the $A$-$B$ model. $A$ lines are solid
whereas $B$ lines are dashed, and both propagators have the functional form 
(\ref{propagator}).}
\label{figAB-1}
\end{figure}

The propagators of both $A$ and $B$ are the same as $i\Delta(x;x')$ given in
expression (\ref{propagator}). The other Feynman rules for the bare action
are the $\lambda A \partial B \partial B$ and $\lambda^2 A^2 \partial B \partial 
B$ vertices. All are depicted in Figure~\ref{figAB-1}.  

\section{Large Logarithms in the Single Field Model}

In this section we calculate 1-loop and 2-loop corrections to a variety of 
quantities in the single field model. We begin with the 1-loop self-mass,
which is then used to compute 1-loop corrections to the plane wave mode
function and the response to a point source. The section closes with a
evaluation of the expectation values of the field and its square at 1-loop 
and 2-loop orders.

\subsection{The Self-Mass}

The 1PI 2-point function can be expressed as,
\begin{equation}
-i M^2(x;x') \equiv \Biggl\langle \Omega \Biggl\vert \frac{i \delta 
S[\Phi]}{\delta \Phi(x)} \frac{i \delta S[\Phi]}{\delta \Phi(x')} + 
\frac{i \delta^2S[\Phi]}{\delta \Phi(x) \delta \Phi(x')} \Biggr\vert
\Omega \Biggr\rangle \; . \label{Mdef}
\end{equation}
Substituting (\ref{varPhi1}) into the first term and using (\ref{PsiofPhi})
gives, 
\begin{eqnarray}
\lefteqn{-i M^2_{\Phi3}(x;x') = \Bigl(\frac{i \lambda}{2}\Bigr)^2 \Biggl\{ 
\frac12 \mathcal{D} \mathcal{D}' \Bigl[ i\Delta(x;x')\Bigr]^2 } \nonumber \\
& & \hspace{-0.5cm} - \mathcal{D} \Bigl[ {a'}^{D-2} {\partial'}^{\rho} 
i\Delta(x;x') \partial'_{\rho} i\Delta(x;x') \Bigr] \!-\! \mathcal{D}' 
\Bigl[ a^{D-2} \partial^{\mu} i\Delta(x;x') \partial_{\mu} i\Delta(x;x') 
\Bigr] \nonumber \\
& & \hspace{4.8cm} + 2 (a a')^{D-2} \partial^{\mu} {\partial'}^{\rho} 
i\Delta(x;x') \partial_{\mu} \partial'_{\rho} i\Delta(x;x') \Biggr\} . 
\qquad \label{MPhi30}
\end{eqnarray}
The second term in (\ref{Mdef}) comes from the second variation 
(\ref{varPhi2}),
\begin{eqnarray}
\lefteqn{-i M^2_{\Phi4}(x;x') = \frac{i \lambda^2}{4} \Biggl\{ -
\delta^D(x \!-\!x') \, a^{D-2} \partial'_{\mu} \partial^{\mu} i\Delta(x;x')
} \nonumber \\
& & \hspace{6cm} + \partial^{\mu} \Bigl[ i\Delta(x;x) a^{D-2} \partial_{\mu} 
\delta^D(x \!-\! x') \Bigr] \Biggr\} . \qquad \label{MPhi40}
\end{eqnarray}

The 3-point contribution (\ref{MPhi30}) can be reduced by a series of 
partial integrations whose general form will occur repeatedly. We will 
present them this once in detail and not again,
\begin{eqnarray}
\lefteqn{ a^{D-2} \partial^{\mu} i\Delta(x;x') \partial_{\mu} i\Delta(x;x')
= \partial^{\mu} \Bigl[ a^{D-2} i\Delta(x;x') \partial_{\mu} 
i\Delta(x;x')\Bigr] } \nonumber \\
& & \hspace{1cm} - i\Delta(x;x') \mathcal{D} i\Delta(x;x') = \frac12 
\mathcal{D} \Bigl[ i\Delta(x;x')\Bigr]^2 \!- i\Delta(x;x) i\delta^D(x 
\!-\! x') \; , \qquad \label{ID1} \\
\lefteqn{2 (a a')^{D-2} \partial^{\mu} {\partial'}^{\rho} i\Delta(x;x')
\partial_{\mu} \partial'_{\rho} i\Delta(x;x') = 2 \partial^{\mu} \Bigl[ 
(a a')^{D-2} {\partial'}^{\rho} i\Delta(x;x') } \nonumber \\
& & \hspace{2.5cm} \times \partial_{\mu} \partial'_{\rho} i\Delta(x;x') 
\Bigr] - 2 {a'}^{D-2} {\partial'}^{\rho} i\Delta(x;x') \partial'_{\rho} 
\mathcal{D} i \Delta(x;x') \; , \qquad \\
& & \hspace{-0.5cm} = \mathcal{D} \Bigl[ {a'}^{D-2} {\partial'}^{\rho}
i\Delta(x;x') \partial'_{\rho} i\Delta(x;x') \Bigr] - 2 {a'}^{D-2} 
{\partial'}^{\rho} i\Delta(x;x') \partial'_{\rho} i\delta^D(x \!-\! x')
\; , \qquad \\
& & \hspace{-0.5cm} = \frac12 \mathcal{D} \mathcal{D}' \Bigl[ 
i\Delta(x;x') \Bigr]^2 \!\!\!-\! \mathcal{D} \Bigl[ i\Delta(x;x) i\delta^D(x 
\!-\! x') \Bigr] \!-\! 2 k H a^{D-1} \partial_0 i\delta^D(x \!-\! x') \; .
\qquad \label{ID2}
\end{eqnarray}
Reducing the four parts of (\ref{MPhi30}) gives,
\begin{equation}
-i M^2_{\Phi3}(x;x') = \frac{\lambda^2}{4} \Biggl\{ -i\Delta(x;x)
\mathcal{D} \Bigl[ i\delta^D(x \!-\! x')\Bigr] + 2 k H a^{D-1} \partial_0
i \delta^D(x \!-\! x') \Biggr\} . \label{MPhi3final}
\end{equation}
Applying the similar reductions to (\ref{MPhi40}) allows the 4-point 
contribution to be expressed as,
\begin{eqnarray}
\lefteqn{-i M^2_{\Phi4}(x;x') = \frac{\lambda^2}{4} \Biggl\{ i\Delta(x;x)
\mathcal{D} \Bigl[ i\delta^D(x \!-\! x')\Bigr] } \nonumber \\
& & \hspace{2.4cm} - 2 k H a^{D-1} \partial_0 i \delta^D(x \!-\! x') + 
(D \!-\! 1) k H^2 a^D i\delta^D(x \!-\! x') \Biggr\} . \qquad 
\label{MPhi4final}
\end{eqnarray}

When (\ref{MPhi3final}) and (\ref{MPhi4final}) are added, all the divergences
cancel and we can take the unregulated limit for the final result,
\begin{equation}
-i M^2_{\Phi}(x;x') = \frac{3 \lambda^2 H^4 a^4}{32 \pi^2} \, 
i \delta^4(x \!-\! x') \; . \label{MPhi}
\end{equation}
Equation (\ref{MPhi}) corresponds to a tachyonic mass of $m^2_{\Phi} = 
-3 \lambda^2 H^4/32\pi^2$. Note that the unit S-matrix implied by 
(\ref{PsiofPhi}) does not preclude interactions from changing the 
free field kinematics. We will see that evolution can also occur, and
that composite operators still require field strength renormalization.

\subsection{1-Loop Mode Function and Exchange Potential}

The self-mass supplies the quantum correction to the linearized effective 
field equation for $\Phi(x)$,
\begin{equation}
\mathcal{D} \Phi(x) - \int \!\! d^4x' M^2(x;x') \Phi(x') = \Bigl[ \mathcal{D} 
+ \frac{3 \lambda^2 H^4 a^4}{32 \pi^2} + O(\lambda^4) \Bigr] \Phi(x) = J(x) 
\; , \label{Phieqn}
\end{equation}
where $\mathcal{D} \equiv \partial^{\mu} a^2 \partial_{\mu}$ is 
the kinetic operator which was introduced in equation (\ref{propeqn}). 
We will study 1-loop corrections to the kinematics of free scalar fields
(with $J(x) = 0$) and to the response to a point source (with $J(x) = K a 
\delta^3(\vec{x})$). It will be useful to consider the scale factor $a$ as 
the time variable,
\begin{equation}
\mathcal{D} \equiv a^2 \Bigl[ -\partial_0^2 - 2 a H \partial_0 + \nabla^2\Bigr]
= a^4 H^2 \Bigl[-a^2 \partial_a^2 - 4 a \partial_a + \frac{\nabla^2}{a^2 H^2}
\Bigr] \; . \label{mathD}
\end{equation}

\subsubsection{Mode Function}

Scalar radiation takes the form,
\begin{equation}
J(x) = 0 \qquad \Longrightarrow \qquad \Phi(x) = u_{\Phi}(\eta,k) 
e^{i \vec{k} \cdot \vec{x}} \; , 
\end{equation}
where the mode function $u_{\Phi}(\eta,k)$ obeys,
\begin{equation}
\Bigl[a^2 \partial_a^2 + 4 a \partial_a + \frac{k^2}{a^2 H^2} - 
\frac{3 \lambda^2 H^2}{32 \pi^2} + O(\lambda^4)\Bigr] u_{\Phi}(\eta,k) 
= 0 \; . \label{uPhieqn}
\end{equation}
The canonically normalized solution for Bunch-Davies vacuum is, 
\begin{equation}
u_{\Phi}(\eta,k) = i \sqrt{\frac{\pi}{4 H a^3}} \, H^{(1)}_{\nu}\Bigl(
\frac{k}{a H}\Bigr) \qquad , \qquad \nu \equiv \sqrt{\frac{9}{4} \!-\!
\frac{m^2_{\Phi}}{H^2}} \; . \label{uPhi}
\end{equation}
The form at late times is,
\begin{equation}
u_{\Phi}(\eta,k) \longrightarrow \frac{\Gamma(\nu)}{\sqrt{4 \pi H a^3}}
\Bigl( \frac{2 a H}{k}\Bigr)^{\nu} \Bigl\{ 1 + \frac1{\nu \!-\! 1}
\Bigl( \frac{k}{2 a H}\Bigr)^2 + \dots \Bigr\} \; . \label{Phiu}
\end{equation}
Because $\nu = \frac32 - \frac{m^2}{3 H^2} + O(\lambda^4)$ is greater than 
$\frac32$ for tachyonic masses, we see that the mode function (\ref{Phiu})
experiences slow growth at late times.

\subsubsection{Exchange Potential}

The exchange potential is the response to a point source,
\begin{equation}
J(x) = K a \delta^3(\vec{x}) \qquad \Longrightarrow \qquad \Phi(x) =
P_{\Phi}(\eta,r) \; , 
\end{equation}
where $r \equiv \Vert \vec{x} \Vert$ and the potential obeys,
\begin{equation}
a^4 H^2 \Bigl[ -a^2 \partial_a^2 - 4 a \partial_a + \frac{\nabla^2}{a^2 H^2}
+ \frac{3 \lambda^2 H^2}{32 \pi^2} + O(\lambda^4)\Bigr] P_{\Phi}(\eta,r)
= K a \delta^3(\vec{x}) \; . \label{PPhieqn}
\end{equation} 
The order $\lambda^0$ solution and its late time limit are \cite{Glavan:2019yfc},
\begin{eqnarray}
P_0(\eta,r) & = & \frac{KH}{4\pi} \Bigl\{ \ln(Hr) + \ln\Bigl(1 \!+\! 
\frac1{a H r}\Bigr) - \frac1{a H r}\Bigr\} \; , \\
& \longrightarrow & \frac{KH}{4\pi} \Bigl\{ \ln(Hr) - \frac1{2 (a H r)^2} + 
\dots \Bigr\} \; . \label{P0}
\end{eqnarray}

The simplest way of solving equation (\ref{PPhieqn}) is by using the retarded
Green's function for a massive, minimally coupled scalar,
\begin{eqnarray}
\lefteqn{m^2 \neq 0 \Longrightarrow G_{\rm ret}(x;x') = -\frac1{4\pi} \Biggl\{
\frac{\delta(\Delta \eta \!-\! \Delta r)}{a a' \Delta r} } \nonumber \\
& & \hspace{2.4cm} - \frac{H^2 \theta(\Delta \eta \!-\! \Delta r)}{2 
\Gamma(\frac12 \!+\! \nu) \Gamma(\frac12 \!-\! \nu)} \sum_{n=0}^{\infty} 
\frac{\Gamma(\frac32 \!+\! \nu \!+\! n) \Gamma(\frac32 \!-\! \nu \!+\! n)}{n! 
(n\!+\!1)!} \Bigl(\frac{y}{4} \Bigr)^n \Biggr\} , \qquad \label{Gretmass}
\end{eqnarray}
where $\nu^2 \equiv \frac94 - \frac{m^2}{H^2}$ and $y \equiv a a' H^2 \Delta x^2$
is the de Sitter length function. (Note that expression (\ref{Gretmass}) takes
the form predicted by DeWitt and Brehme \cite{DeWitt:1960fc} with a universal
light-cone singularity plus a mass-dependent tail term.) For small $m^2/H^2$
we can expand $G_{\rm ret}(x;x')$,
\begin{eqnarray}
\lefteqn{ G_{\rm ret}(x;x') = -\frac1{4\pi} \Biggl\{ \frac{\delta(\Delta \eta
\!-\! \Delta r)}{a a' \Delta r} + H^2 \theta(\Delta \eta \!-\! \Delta r) }
\nonumber \\
& & \hspace{2.7cm} - m^2 \theta(\Delta \eta \!-\! \Delta r) \Biggl[ \frac13 
\ln\Bigl( 1 \!-\! \frac{y}{4}\Bigr) \!+\! \frac12 \!+\! 
\frac{\frac{y}{6}}{y \!-\! 4} \Biggr] + O(m^4) \Biggr\} . \qquad 
\label{Gretexp}
\end{eqnarray}
Integrating against the source and taking the late time limit gives,
\begin{eqnarray}
P_{\Phi}(\eta,r) &\!\!\! = \!\!\!& \int \!\! d^4x' G_{\rm ret}(x;x') \!\times\! 
K a' \delta^3(\vec{x}') \; , \\
&\!\!\! \longrightarrow \!\!\!& \frac{K H}{4 \pi} \Biggl\{ \ln(Hr) +
\frac{\lambda^2 H^2}{32 \pi^2} \ln(a) \ln(Hr) + O(\lambda^4) \Biggr\} \; .
\label{PhiP}
\end{eqnarray} 

\subsection{The Expectation Values of $\Phi(x)$ and $\Phi^2(x)$}

The field definition (\ref{PsiofPhi}) makes it simple to evaluate expectation
values of $\Phi(x)$ and its square. The first power is,
\begin{equation}
\Phi[\Psi] =  \Psi - \frac14 \lambda \Psi^2 + \frac18 \lambda^2 \Psi^3 -
\frac{5}{24} \lambda^3 \Psi^4 + O\Bigl( \lambda^4 \Psi^5\Bigr) \; . 
\label{Phi1}
\end{equation}
Taking the expectation value gives,
\begin{equation}
\Bigl\langle \Omega \Bigl\vert \Phi(x) \Bigr\vert \Omega \Bigr\rangle = 
-\frac14 \lambda i\Delta(x;x) - \frac{15}{64} \lambda^3 \Bigl[ i\Delta(x;x)
\Bigr]^2 + O\Bigl( \lambda^5 \Bigl[ i\Delta(x;x)\Bigr]^3 \Bigr) \; . 
\label{PhiVEV}
\end{equation}
Expression (\ref{coincprop}) shows that the coincident propagator is time 
dependent, so we see that $\langle \Omega \vert \Phi(x) \vert \Omega \rangle$ 
evolves, in spite of vanishing flat space scattering amplitudes.

The expansion of $\Phi^2$ is,
\begin{equation}
\Phi^2(x) = \Psi^2(x) - \frac12 \lambda \Psi^3(x) + \frac{5}{16} \lambda^2
\Psi^4(x) + O\Bigl( \lambda^3 \Psi^5\Bigr) \; . \label{Phisq}
\end{equation}
Taking its expectation value gives, 
\begin{equation}
\Bigl\langle \Omega \Bigl\vert \Phi^2(x) \Bigr\vert \Omega \Bigr\rangle =
i\Delta(x;x) + \frac{15}{16} \lambda^2 \Bigl[ i\Delta(x;x)\Bigr]^2 + 
O\Bigl( \lambda^4 \Bigl[ i\Delta(x;x)\Bigr]^3 \Bigr) \; . \label{PhisqVEV}
\end{equation}
$\Phi^2(x)$ is a composite operator and requires renormalization with 
counterterms of the form,
\begin{equation}
\delta \Phi^2 = K_{\Phi 1} R + K_{\Phi 2} R \Phi^2 + K_{\Phi 3} R^2 + 
O(\lambda^4) \; . \label{deltaPhi2}
\end{equation}
Comparison with the primitive expression (\ref{PhisqVEV}) implies,
\begin{eqnarray}
K_{\Phi 1} &\!\!\! = \!\!\!& \frac{\mu^{D-4}}{(4\pi)^{\frac{D}2}} 
\frac{\Gamma(D\!-\!1)}{\Gamma(\frac{D}2)} \frac{\pi {\rm cot}(
\frac{D\pi}{2})}{D (D\!-\!1)} \; , \\
K_{\Phi 2} &\!\!\! = \!\!\!& \frac{15 \lambda^2 \mu^{D-4}}{8 
(4\pi)^{\frac{D}2}} \frac{\Gamma(D\!-\!1)}{\Gamma(\frac{D}2)} 
\frac{\pi {\rm cot}(\frac{D\pi}{2})}{D (D\!-\!1)} \; , \\
K_{\Phi 3} &\!\!\! = \!\!\!& \frac{15 \lambda^2 \mu^{D-4}}{16
(4\pi)^{D}} \Bigl[\frac{\Gamma(D\!-\!1)}{\Gamma(\frac{D}2)} 
\frac{\pi {\rm cot}(\frac{D\pi}{2})}{D (D\!-\!1)} \Biggr]^2 \; . 
\end{eqnarray}
Using these values in (\ref{deltaPhi2}), adding $\delta \Phi^2$ to
(\ref{PhisqVEV}), and taking the unregulated limit gives the fully 
renormalized result,
\begin{equation}
\Bigl\langle \Omega \Bigl\vert \Phi^2(x) \Bigr\vert \Omega 
\Bigr\rangle_{\rm ren} = \frac{H^2}{4 \pi^2} \ln\Bigl(\frac{\mu a}{H}
\Bigr) + \frac{15 \lambda^2 H^4}{256 \pi^4} \ln^2\Bigl( \frac{\mu a}{H}
\Bigr) + O(\lambda^4) \; . \label{Phi2ren}
\end{equation}

\section{Large Logarithms in the Two Field Model}

The task of this section is computing the same things for the two 
field model (\ref{LAB}) that we previously did for the single field 
model (\ref{LPhi}). The order of presentation is the same as in the
previous section, although our labor is complicated by the inability
to remove interactions by a local field redefinition. We must also 
digress to explain the Schwinger-Keldysh formalism when solving the
effective field equations.

\subsection{1-Loop Self-Masses for $A$ and $B$}

The four counterterms we require for renormalizing the self-masses at 1-loop are,
\begin{eqnarray}
\lefteqn{\Delta \mathcal{L} = -\frac12 C_{A1} \square A \square A \sqrt{-g} - 
\frac12 C_{A2} \, R \partial_{\mu} A \partial_{\nu} A g^{\mu\nu} \sqrt{-g} } 
\nonumber \\
& & \hspace{3.5cm} - \frac12 C_{B1} \square B \square B \sqrt{-g} - \frac12 C_{B2} 
\, R \partial_{\mu} B \partial_{\nu} B g^{\mu\nu} \sqrt{-g} \; . \qquad 
\label{Cterms}
\end{eqnarray}

\begin{figure}[H]
\centering
\includegraphics[width=11cm]{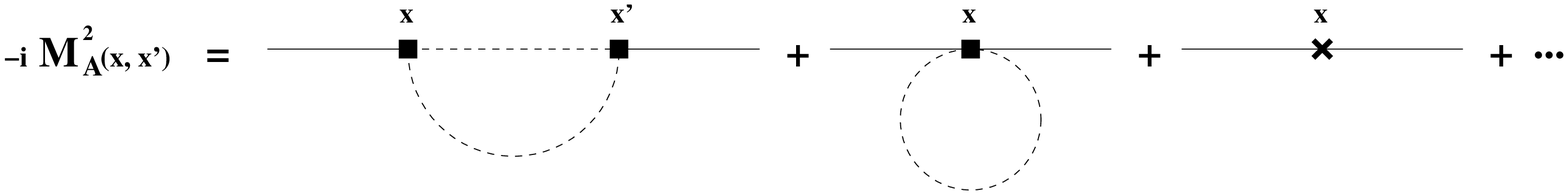}
\caption{\footnotesize Diagrams which represent the 1-loop contributions to 
the $A$-field self-mass $-i M^2_{A}(x;x')$. Recall that $A$ lines are solid, 
whereas $B$ lines are dashed.}
\label{figAB-Ma}
\end{figure}

The three diagrams which contribute to the $A$ self-mass at 1-loop are shown
in Figure~\ref{figAB-Ma}.
\noindent From left to right, their analytic expressions are,
\begin{eqnarray}
-i M^2_{A3}(x;x') &\!\!\! = \!\!\!& \frac{(-i \lambda)^2}{2} (a a')^{D-2} 
\partial^{\mu} {\partial'}^{\rho} i\Delta(x;x') \partial_{\mu} \partial'_{\rho} 
i\Delta(x;x') \; , \qquad \label{MA30} \\
-i M^2_{A4}(x;x') &\!\!\! = \!\!\!& -\frac{\lambda^2}{4} i\delta^D(x \!-\! x')
\, a^{D-2} \partial^{\mu} \partial'_{\mu} i\Delta(x;x') \; , \qquad 
\label{MA40} \\
-i M^2_{Ac}(x;x') &\!\!\! = \!\!\!& -C_{A1} \mathcal{D} \mathcal{D}' \Bigl[
\frac{i \delta^D(x \!-\! x')}{(a a')^{\frac{D}2}} \Bigr] + C_{A2} \partial^{\mu}
\Bigl[ R a^{D-2} \partial_{\mu} i\delta^D(x \!-\! x')\Bigr] \; . \qquad 
\label{MAc0}
\end{eqnarray}

After the same sort of reductions employed in the single-field model, the
two primitive diagrams take the forms, 
\begin{eqnarray}
-i M^2_{A3}(x;x') &\!\!\! = \!\!\!& \frac{(-i\lambda)^2}{2} \Biggl\{ \frac14 
\mathcal{D} \mathcal{D}' \Bigl[ i\Delta(x;x')\Bigr]^2 - \frac12 \mathcal{D}
\Bigl[i \Delta(x;x) i\delta^D(x \!-\! x') \Bigr] \nonumber \\
& & \hspace{5.2cm} - k H a^{D-1} \partial_0 i \delta^D(x \!-\! x') \Biggr\} , 
\qquad \label{MA31} \\
-i M^2_{A4}(x;x') &\!\!\! = \!\!\!& -\frac{i \lambda^2}{4} \delta^D(x \!-\! x')
\!\times\! -(D-1) k H^2 a^D \; . \qquad \label{MA41} 
\end{eqnarray}
The square of the propagator in expression (\ref{MA31}) is logarithmically 
divergent so we need only retain dimensional regularization for the leading
term and can take $D=4$ for the rest,
\begin{eqnarray}
\lefteqn{ \Bigl[ i\Delta(x;x')\Bigr]^2 = \frac{\Gamma^2(\frac{D}2 \!-\! 1)}{
16 \pi^D} \frac1{(a a' \Delta x^2)^{D-2}} } \nonumber \\
& & \hspace{2.2cm} - \frac{H^2}{16 \pi^4} \frac{\ln(\frac14 H^2 \Delta x^2)}{
a a' \Delta x^2} + \frac{H^4}{64 \pi^4} \ln^2\Bigl(\frac14 H^2 \Delta x^2\Bigr)
+ O(D \!-\! 4) \; . \qquad \label{propsquare}
\end{eqnarray} 
The fundamental logarithmic divergence is $1/\Delta x^{2D-4}$. We localize this
by first extracting a d`Alembertian, then adding zero in the form of the massless 
propagator equation in flat space \cite{Onemli:2002hr,Onemli:2004mb},
\begin{eqnarray}
\lefteqn{\frac1{\Delta x^{2D-4}} = \frac{\partial^2}{2 (D \!-\! 3) (D\!-\!4)} 
\Bigl[ \frac1{\Delta x^{2D-6}}\Bigr] = \frac{\mu^{D-4}}{2 (D\!-\! 3) (D\!-\!4)} 
\frac{4 \pi^{\frac{D}2} i \delta^D(x \!-\! x')}{\Gamma( \frac{D}2 \!-\! 1)} } 
\nonumber \\
& & \hspace{5cm} + \frac{\partial^2}{2 (D\!-\!3) (D\!-\!4)} \Bigl[ 
\frac1{\Delta x^{2D-6}} - \frac{\mu^{D-4}}{\Delta x^{D-2}}\Bigr] \; , \qquad \\
& & \hspace{0.4cm} = \frac{\mu^{D-4}}{2 (D\!-\! 3) (D\!-\!4)} \frac{4 \pi^{\frac{D}2} 
i \delta^D(x \!-\! x')}{\Gamma( \frac{D}2 \!-\! 1)} -\frac{\partial^2}{4} 
\Bigl[ \frac{\ln(\mu^2 \Delta x^2)}{\Delta x^2}\Bigr] + O(D\!-\!4) \; . \qquad 
\end{eqnarray}
Here $\mu$ is the mass scale of dimensional regularization.

Comparison with expression (\ref{MAc0}) gives the two $A$-type counterterms,
\begin{equation}
C_{A1} = -\frac{\lambda^2 \mu^{D-4}}{32 \pi^{\frac{D}2}}
\frac{\Gamma(\frac{D}2 \!-\! 1)}{2 (D\!-\!3) (D\!-\!4)} \quad , \quad 
C_{A2} = \frac{\lambda^2 \mu^{D-4}}{4 (4\pi)^{\frac{D}2}} 
\frac{\Gamma(D\!-\!1)}{\Gamma(\frac{D}2)} \frac{\pi {\rm cot}(\frac{D \pi}{2})}{
D (D \!-\! 1)} \; . \label{CA12}
\end{equation}
Combining $-iM^2_{A3}(x;x')$ and $-iM^2_{A4}(x;x')$ with $-i M^2_{3c}(x;x')$
and taking the unregulated limit gives the renormalized self-mass at one
loop,  
\begin{eqnarray}
\lefteqn{-i M^2_{A}(x;x') = -\frac{3 \lambda^2 H^4 a^4}{32 \pi^2} \, 
i\delta^4(x \!-\!x') + \frac{\lambda^2 H^2 \partial^{\mu}}{16 \pi^2} 
\Bigl[\ln\Bigl( \frac{\mu a}{H} \Bigr) a^2 \partial_{\mu} i\delta^4(x 
\!-\! x') \Bigr] } \nonumber \\
& & \hspace{2cm} + \frac{\lambda^2 \mathcal{D} \mathcal{D}'}{512 \pi^4}
\Biggl\{ \frac{ \ln(a a') 4\pi^2 i \delta^4(x \!-\! x')}{(a a')^2} + 
\frac{\partial^2}{(a a')^2} \Bigl[ \frac{\ln(\mu^2 \Delta x^2)}{
\Delta x^2}\Bigr] \nonumber \\
& & \hspace{4.5cm} + \frac{4 H^2}{a a'} \frac{\ln(\frac14 H^2 
\Delta x^2)}{\Delta x^2} - H^4 \ln^2\Bigl( \frac14 H^2 \Delta x^2\Bigr)
\Biggr\} . \qquad \label{MAren}
\end{eqnarray}
Note that the first term represents a positive mass-squared of the same 
magnitude as the tachyonic mass we saw in expression (\ref{MPhi}). It is
accompanied by many other contributions which signal that this system
is not reducible to a free field.

\begin{figure}[H]
\centering
\includegraphics[width=11cm]{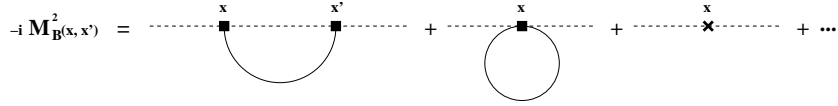}
\caption{\footnotesize Diagrams which represent the 1-loop contributions to 
the $B$-field self-mass $-i M^2_{B}(x;x')$. Recall that $A$ lines are solid, 
whereas $B$ lines are dashed.}
\label{figAB-Mb}
\end{figure}

Figure~\ref{figAB-Mb} depicts the three diagrams which contribute to 
the self-mass of $B$ at 1-loop. 
\noindent The analytic expressions for these three diagrams are, from left
to right,
\begin{eqnarray}
-i M^2_{B3}(x;x') &\!\!\! = \!\!\!& (i \lambda)^2 \partial^{\mu} 
{\partial'}^{\rho} \Bigl[ i\Delta(x;x') (a a')^{D-2} \partial_{\mu} 
\partial'_{\rho} i\Delta(x;x') \Bigr] \; , \qquad \label{MB30} \\
-i M^2_{B4}(x;x') &\!\!\! = \!\!\!& \frac{\lambda^2}{4} \partial^{\mu} \Bigl[
i\Delta(x;x) a^{D-2} \partial_{\mu} i\delta^D(x \!-\! x') \Bigr] \; , \qquad
\label{MB40} \\
-i M^2_{Bc}(x;x') &\!\!\! = \!\!\!& -C_{B1} \mathcal{D} \mathcal{D}' \Bigl[
\frac{i \delta^D(x \!-\! x')}{(a a')^{\frac{D}2}} \Bigr] + C_{B2} \partial^{\mu}
\Bigl[ R a^{D-2} \partial_{\mu} i\delta^D(x \!-\! x')\Bigr] \; . \qquad 
\label{MBc0}
\end{eqnarray}

After some familiar reductions the two primitive diagrams take the form,
\begin{eqnarray}
-i M^2_{B3}(x;x') &\!\!\! = \!\!\!& \frac{(i \lambda)^2}{2} \mathcal{D} 
\mathcal{D}' \Bigl[ i\Delta(x;x')\Bigr]^2 \nonumber \\
& & \hspace{1.5cm} - (i\lambda)^2 \partial^{\mu} {\partial'}^{\rho} \Bigl[ 
(a a')^{D-2} \partial_{\mu} i\Delta(x;x') \partial'_{\rho} i\Delta(x;x')
\Bigr] \; , \qquad \label{MB31} \\
-i M^2_{B4}(x;x') &\!\!\! = \!\!\!& -\frac{\lambda^2}{4} k \pi {\rm cot}\Bigl(
\frac{D \pi}{2}\Bigr) \mathcal{D} \Bigl[i\delta^D(x \!-\! x') \Bigr] 
\nonumber \\
& & \hspace{1.8cm} + \frac{\lambda^2 H^2 \partial^{\mu}}{16 \pi^2} \Bigl[ 
\ln(a) a^2 \partial_{\mu} i \delta^4(x \!-\! x')\Bigr] + O(D \!-\! 4) \; . 
\qquad \label{MB41}
\end{eqnarray}
Expression (\ref{MB41}) is already fully reduced, and the reduction of the 
first term of (\ref{MB31}) is identical to that of (\ref{propsquare}), but 
the second term requires new analysis. Because this contribution is 
quadratically divergent we must retain dimensional regularization for the 
first two terms in the power series of the propagator,
\begin{equation}
\partial_{\mu} i\Delta(x;x') = -\frac{\Gamma(\frac{D}2)}{2 \pi^{\frac{D}2}
(a a')^{\frac{D}2 - 1}} \Biggl\{ \frac{\Delta x_{\mu}}{\Delta x^D} +
\frac{[\frac12 a H \delta^0_{~\mu} \!+\! \frac{D}{8} a a' H^2 \Delta 
x_{\mu}]}{\Delta x^{D-2}} + \dots \Biggr\} . \label{dprop}
\end{equation}
Taking the product of two such differentiated propagators, extracting 
derivatives and taking $D=4$ in integrable terms gives,
\begin{eqnarray}
\lefteqn{-(i\lambda)^2 \partial^{\mu} {\partial'}^{\rho} \Bigl[ (a a')^{D-2}
\partial_{\mu} i\Delta(x;x') \partial'_{\rho} i\Delta(x;x') \Bigr] } 
\nonumber \\
& & \hspace{0cm} = - \frac{(i\lambda)^2 \Gamma^2(\frac{D}2)}{4 \pi^D} 
\Biggl\{ \frac{\mathcal{D} \mathcal{D}'}{4 (D \!-\! 2)^2} \Bigl[ 
\frac1{(a a' \Delta x^2)^{D-2}}\Bigr] - \frac{D H^2 \partial \!\cdot\! 
\partial'}{8 (D \!-\! 2)} \Bigl[ \frac{a a'}{\Delta x^{2 D - 4}}\Bigr] 
\nonumber \\
& & \hspace{1cm} + \frac18 \mathcal{D} \mathcal{D}' \Bigl[ \frac{H^2}{a a' 
\Delta x^2} \Bigr] \!+\! \frac{a^2 {a'}^2 H^4 \nabla^2}{8} \Bigl[ 
\frac1{\Delta x^2}\Bigr] \!-\! \frac{H^4}{16} \mathcal{D} \mathcal{D}' 
\ln(H^2 \Delta x^2) \Biggr\} . \qquad \label{2ndterm}
\end{eqnarray}

Comparison with expression (\ref{MBc0}) gives the two $B$-type counterterms,
\begin{eqnarray}
C_{B1} &\!\!\! = \!\!\!& -\frac{\lambda^2 \mu^{D-4}}{16 \pi^{\frac{D}2}}
\frac{\Gamma(\frac{D}2 \!-\! 1)}{2 (D\!-\!3) (D\!-\!4)} \; , \label{CB1} \\
C_{B2} &\!\!\! = \!\!\!& \frac{\lambda^2 \mu^{D-4}}{4 (4\pi)^{\frac{D}2}} 
\frac{\Gamma(D\!-\!1)}{\Gamma(\frac{D}2)} \frac{\pi {\rm cot}(\frac{D \pi}{2})}{
D (D \!-\! 1)} - \frac{\lambda^2 \mu^{D-4}}{32 \pi^{\frac{D}2}} 
\frac{\Gamma(\frac{D}2 \!-\! 1)}{2 (D\!-\!3) (D\!-\!4)} 
\Bigl(\frac{D\!-\!2}{D\!-\!1}\Bigr) \; . \qquad \label{CB2}
\end{eqnarray}
Note that $C_{B2}$ cancels divergences in $-i M^2_{B4}(x;x')$ --- the left 
hand contribution to (\ref{CB2}) --- and in $-i M^2_{B3}(x;x')$ --- the
right hand term of (\ref{CB2}). Combining the two primitive diagrams with the
counterterm and taking the unregulated limit gives,
\begin{eqnarray}
\lefteqn{-i M^2_{B}(x;x') = \frac{\lambda^2 \mathcal{D} \mathcal{D}'}{
64 \pi^2} \Bigl[ \frac{ \ln(a a') i \delta^4(x \!-\! x')}{(a a')^2} \Bigr] 
\!-\! \frac{\lambda^2 H^2 \partial^{\mu}}{16 \pi^2} \Bigl[ 
\ln\Bigl(\! \frac{H a}{\mu} \! \Bigr) a^2 \partial_{\mu} i\delta^4(x \!-\! x') 
\Bigr] } \nonumber \\
& & \hspace{-0.6cm} + \frac{\lambda^2 \mathcal{D} \mathcal{D}'}{256 \pi^4}
\Biggl\{ \frac{\partial^2}{(a a')^2} \Bigl[ \frac{\ln(\mu^2 \Delta x^2)}{
\Delta x^2}\Bigr] + \frac{H^2 \partial^2}{a a'} \Bigl[ \ln^2\Bigl(
\frac{ H^2 \Delta x^2}{4}\Bigr)\Bigr] - 2 H^4 \Bigl[ \ln^2\Bigl( 
\frac{H^2 \Delta x^2}{4} \Bigr) \nonumber \\
& & \hspace{-0.6cm} + 2 \ln\Bigl(\!\frac{H^2 \Delta x^2}{4}\!\Bigr) \!\Bigr] 
\!\!\Biggr\} \!+\! \frac{\lambda^2 H^2 \partial \!\cdot\! \partial'}{64 \pi^4} 
\Biggl\{ \!a a' \partial^2 \Bigl[ \frac{\ln(\mu^2 \Delta x^2)}{\Delta x^2}
\Bigr] \!\!\Biggr\} \!+\! \frac{\lambda^2 H^4 (a a')^2 \nabla^2}{32 \pi^4} 
\frac1{\Delta x^2} . \qquad \label{MBren} 
\end{eqnarray}

\subsection{1-Loop Mode Functions \& Exchange Potentials}

\subsubsection{Schwinger-Keldysh Effective Field Equations}

The linearized effective field equations for $A$ reads,
\begin{equation}
\mathcal{D} A(x) = J(x) + \int \!\! d^4x' \, M^2_{A}(x;x') A(x') \; .
\label{effective}
\end{equation}
The $B$ equation is the same with $A(x)$ replaced by $B(x)$ and $M^2_{A}(x;x')$
replaced by $M^2_{B}(x;x')$. Setting the source $J(x) = 0$ describes the 
propagation of scalar radiation, while the choice $J(x) = a(\eta) \delta^3(\vec{x})$ 
gives the scalar exchange potential.

Using our in-out results (\ref{MAren}) and (\ref{MBren}) in equation 
(\ref{effective}) would result in two problems:
\begin{itemize}
\item{The fact that the self-masses are complex precludes real solutions; and}
\item{The fact that the self-masses $M^2_{A,B}(x;x')$ are nonzero for ${x'}^{\mu}$
outside the past light-cone of $x^{\mu}$ leads to a response {\it before} its
cause.}
\end{itemize}
Both of these embarrassments can be avoided by employing the self-masses of
the Schwinger-Keldysh formalism \cite{Schwinger:1960qe,Mahanthappa:1962ex,
Bakshi:1962dv,Bakshi:1963bn}. There are many fine references on the 
Schwinger-Keldysh effective field equations \cite{Chou:1984es,Jordan:1986ug,
Calzetta:1986ey} but we only need the simple rules for converting the in-out 
self-masses (\ref{MAren}) and (\ref{MBren}) to in-in self-masses 
\cite{Ford:2004wc}:
\begin{itemize}
\item{The Dirac delta function terms are not changed;}
\item{For every term involving the Poincar\'e interval $\Delta x^2(x;x')$,
defined in expression (\ref{Dx2def}), one must subtract the very same 
function of,
\begin{equation}
\Delta x^2_{+-}(x;x') \equiv \Bigl\Vert \vec{x} - \vec{x}' \Bigr\Vert^2
- \Bigl( \eta - \eta' + i \epsilon\Bigr)^2 \; . \label{Dx2+-}
\end{equation}}
\end{itemize}

In converting the in-out self-mass to its in-in cognate it is desirable 
to extract d`Alembertians from inverse powers of $1/\Delta x^2$ to reach 
powers of logarithms,
\begin{equation}
\frac1{\Delta x^2} = \frac{\partial^2}{4} \Bigl[ \ln(\mu^2 \Delta x^2)\Bigr] 
\quad , \quad \frac{\ln(\mu^2 \Delta x^2)}{\Delta x^2} = \frac{\partial^2}{8} 
\Bigl[ \ln^2(\mu^2 \Delta x^2) \!-\! 2 \ln(\mu^2 \Delta x^2)\Bigr] \; . 
\label{rule2}
\end{equation}
Differences of powers of logarithms of $\Delta x^2$ and $\Delta x^2_{+-}$
give a form that makes the reality and causality of $-i M^2_{A,B}(x;x')$ 
manifest,
\begin{eqnarray}
\ln(\mu^2 \Delta x^2) - \ln(\mu^2 \Delta x^2_{+-}) &\!\!\! = \!\!\!& 
2\pi i \theta(\Delta \eta \!-\! \Delta r) \; , \label{log1} \\
\ln^2(\mu^2 \Delta x^2) - \ln^2(\mu^2 \Delta x^2_{+-}) &\!\!\! = \!\!\!& 
4\pi i \theta(\Delta \eta \!-\! \Delta r) \ln[\mu^2 (\Delta \eta^2 \!-\!
\Delta r^2)] \; . \label{log2}
\end{eqnarray}
where $\Delta \eta \equiv \eta - \eta'$ and $\Delta r \equiv \Vert \vec{x} 
- \vec{x}'\Vert$.

Finally, we must adapt (\ref{effective}) to the fact that only one
loop results for the self-masses are available. This means we can only 
solve the equation perturbatively in powers of $\lambda^2$,
\begin{equation}
A(x) \equiv A_{0}(x) + \lambda^2 A_{1}(x) + O(\lambda^4) \quad , \quad
B(x) \equiv B_0(x) + \lambda^2 B_{1}(x) + O(\lambda^4) \; .
\end{equation}
The zeroth order solutions obey,
\begin{equation}
\mathcal{D} A_0(x) = J_A(x) \quad , \quad \mathcal{D} B_0(x) = J_B(x) 
\; . \label{orderzero}
\end{equation}
At 1-loop order we have,
\begin{eqnarray}
\lefteqn{ \mathcal{D} A_1(x) = \frac{3 H^4 a^4 A_0}{32 \pi^2} - \frac{H^2
\partial^{\mu}}{16 \pi^2} \Bigl[ \ln\Bigl(\frac{\mu a}{H}\Bigr) a^2 
\partial_{\mu} A_0\Bigr] - \frac{\mathcal{D}}{64 \pi^2} \Bigl[ \frac{\ln(a) 
\mathcal{D} A_0}{a^4}\Bigr] } \nonumber \\
& & \hspace{-0.5cm} - \frac{\mathcal{D}}{512 \pi^3} \int \!\! d^4x' \Biggl\{ 
\frac{\partial^4}{2 a^2 {a'}^2} \Biggl[ \theta(\Delta \eta \!-\! \Delta r) 
\Bigl( \ln[\mu^2 (\Delta \eta^2 \!-\! \Delta r^2)] \!-\! 1 \Bigr) \Biggr] 
\nonumber \\
& & \hspace{1.5cm} + \frac{2 H^2 \partial^2}{a a'} \Biggl[ \theta(\Delta \eta
\!-\! \Delta r) \Bigl(\ln\Bigl[\frac14 H^2 (\Delta \eta^2 \!-\! \Delta r^2)
\Bigr] \!-\! 1\Bigr) \Biggr] \nonumber \\
& & \hspace{3cm} - 4 H^4 \theta(\Delta \eta \!-\! \Delta r) \ln\Bigl[\frac14 
H^2 (\Delta \eta^2 \!-\! \Delta r^2)\Bigr] \Biggr\} \mathcal{D}' A_0(x') \; , 
\qquad \label{A1eqn} \\
\lefteqn{ \mathcal{D} B_1(x) = -\frac{\mathcal{D}}{32 \pi^2} 
\Bigl[ \frac{\ln(a) \mathcal{D} B_0}{a^4} \Bigr] + \frac{H^2\partial^{\mu}}{
16 \pi^2} \Bigl[ \ln\Bigl(\frac{H a}{\mu}\Bigr) a^2 \partial_{\mu} B_0\Bigr] 
} \nonumber \\
& & \hspace{-0.5cm} - \frac{\mathcal{D}}{256 \pi^3} \int \!\!
d^4x' \Biggl\{ \frac{\partial^4}{2 a^2 {a'}^2} \Biggl[ \theta(\Delta \eta
\!-\! \Delta r) \Bigl( \ln[\mu^2 (\Delta \eta^2 \!-\! \Delta r^2)] \!-\! 1
\Bigr) \Biggr] \nonumber \\
& & \hspace{2cm} + \frac{4 H^2 \partial^2}{a a'} \Bigl( \theta(\Delta \eta
\!-\! \Delta r) \ln\Bigl[\frac14 H^2 (\Delta \eta^2 \!-\! \Delta r^2)\Bigr]
\Bigr) \nonumber \\
& & \hspace{3cm} - 8 H^4 \theta(\Delta \eta \!-\! \Delta r) \Bigl(
\ln\Bigl[\frac14 H^2 (\Delta \eta^2 \!-\! \Delta r^2)\Bigr] \!+\! 1\Bigr) 
\Biggr\} \mathcal{D}' B_0(x') \nonumber \\
& & \hspace{-0.5cm} -\frac{H^2}{128 \pi^3} \int \!\! d^4x' 
\partial \!\cdot\! \partial' \Biggl\{ a a' \partial^4 \Biggl[
\theta(\Delta \eta \!-\! \Delta r) \Bigl[ \ln[\mu^2 (\Delta \eta^2 \!-\!
\Delta x^2)] \!-\! 1\Bigr] \Biggr] \Biggr\} B_0(x') \nonumber \\
& & \hspace{4cm} -\frac{H^4}{64 \pi^3} \int \!\! d^4x' a^2
{a'}^2 \nabla^2 \partial^2 \Bigl[ \theta(\Delta \eta \!-\! \Delta r)\Bigr]
B_0(x') \; . \qquad \label{B1eqn}
\end{eqnarray}

\subsubsection{1-Loop Corrected Mode Functions}

Scalar radiation corresponds to $J_{A,B}(x) = 0$ and has zeroth order 
solution,
\begin{equation}
A_0(x) = B_0(x) = u_0(\eta,k) e^{i \vec{k} \cdot \vec{x}} \qquad 
\Longrightarrow \qquad u_0(\eta,k) = \frac{H}{\sqrt{2 k^3}} \Bigl[1 
+ i k \eta\Bigr] e^{-i k \eta} \; . \label{zerothrad}
\end{equation}
Because $\mathcal{D} A_0(x) = 0 = \mathcal{D} B_0(x)$, very few terms
of the 1-loop field equations (\ref{A1eqn}-\ref{B1eqn}) survive,
\begin{eqnarray}
\mathcal{D} A_1 &\!\!\! = \!\!\!& \frac{3 H^4 a^4 A_0}{32 \pi^2}
+ \frac{H^3 a^3 \partial_0 A_0}{16 \pi^2} \; , \label{Amode1} \\
\mathcal{D} B_1 &\!\!\! = \!\!\!& -\frac{H^3 a^3 \partial_0 B_0}{16 \pi^2} 
+ \frac{H^4 k^2 a^2}{64 \pi^3} \! \int \!\! d^4x' \, \theta(\Delta \eta
\!-\! \Delta r) {\partial'}^2 \Bigl[ {a'}^2 B_0(x')\Bigr] \noindent \\
& & \hspace{-1.5cm} - \frac{H^2 \partial^{\mu}}{128 \pi^3} \Biggl\{ a
\partial^2 \!\!\! \int \!\! d^4x' \, \theta(\Delta \eta \!-\! \Delta r) 
\Bigl[ \ln[\mu^2 (\Delta \eta^2 \!-\! \Delta r^2)] \!-\! 1\Bigr]
{\partial'}^2 \Bigl[ a' \partial'_{\mu} B_0(x')\Bigr] \Biggr\} . \qquad 
\label{Bmode1}
\end{eqnarray}

1-loop corrections to the $A$ mode function are dominated by the
first term on the right hand side of expression (\ref{Amode1}) which
corresponds to a mass,
\begin{equation}
m^2_{A} = \frac{3 \lambda^2 H^4}{32 \pi^2} + O(\lambda^4) \quad
\Longrightarrow \quad \nu \equiv \sqrt{\frac94 \!-\! \frac{m_A^2}{H^2}}
= \frac32 - \frac{\lambda^2 H^2}{32 \pi^2} + O(\lambda^4) \; . \label{mA2}
\end{equation}
Substituting (\ref{mA2}) into our previous result (\ref{Phiu})
for the late time form of a massive scalar mode function, and expanding 
in powers of $\lambda^2$ gives,
\begin{eqnarray}
\lefteqn{i \sqrt{\frac{\pi}{4 H a^3}} \, H^{(1)}_{\nu}\Bigl(
\frac{k}{a H}\Bigr) \longrightarrow \frac{H}{\sqrt{2 k^3}} \Biggl\{
1 + \frac{k^2}{2 a^2 H^2} + O\Bigl( \frac{k^4}{a^4 H^4} \Bigr) 
-\frac{\lambda^2 H^2}{32 \pi^2} \Biggl[\ln\Bigl(\frac{2 a H}{k}\Bigr) }
\nonumber \\
& & \hspace{0.3cm} + \psi\Bigl(\frac32\Bigr) + \Bigl[\ln\Bigl(\frac{2 a H}{k}
\Bigr) \!+\! \psi\Bigl(\frac32\Bigr) \!-\! 2\Bigr] \frac{k^2}{2 a^2 H^2} + 
O\Bigr(\frac{k^4}{a^4 H^4}\Bigr) \Biggr] + O(\lambda^4) \Biggr\} . \qquad 
\label{Au}
\end{eqnarray}
In contrast, the final term in (\ref{Amode1}) is down by two factors of $a$,
\begin{equation}
\partial_0 u_0(\eta,k) = \frac{H}{\sqrt{2 k^3}} \!\times\! -\frac{k^2}{a H}
\exp\Bigl[\frac{i k}{a H}\Bigr] \; ,
\end{equation}
and corrects the $A$ mode function by terms which fall off like $1/a^2$.

The first term on the right hand side of (\ref{Bmode1}) is opposite to the
final term of (\ref{Amode1}), and is similarly irrelevant. To evaluate the
nonlocal contribution to (\ref{Bmode1}) we first note,
\begin{equation}
{\partial'}^2 \Bigl[ {a'}^2 B_0(x')\Bigr] = 12 {a'}^4 H^2 B_0(x') + 
2 {a'}^3 H \partial_0' B_0(x') \longrightarrow 12 {a'}^4 H^2 \!\times\!
\frac{H}{\sqrt{2 k^3}} \; .
\end{equation}
Hence the leading late time form of the right hand side of (\ref{Bmode1}) is,
\begin{eqnarray}
\lefteqn{ \frac{H^4 k^2 a^2}{64 \pi^3} \int \!\! d^4x' \theta(\Delta \eta 
\!-\! \Delta r) {\partial'}^2 \Bigl[ {a'}^2 B_0(x')\Bigr] } \nonumber \\
& & \hspace{1.5cm} \longrightarrow \frac{H^4 k^2 a^2}{64 \pi^3} \int \!\! 
d^4x' \theta(\Delta \eta \!-\! \Delta r) \!\times\! 12 {a'}^4 H^2 u_0(0,k) 
\; , \\
& & \hspace{1.5cm} = \frac{H^6 k^2 a^2}{4 \pi^2} \, u_0(0,k) \!\! 
\int_{\eta_i}^{\eta} \!\! d\eta' {a'}^4 \Delta \eta^3 \longrightarrow 
\frac{H^2 k^2 a^2 \ln(a)}{4 \pi^2} \, u_0(0,k) \; . \qquad \label{Bsource}
\end{eqnarray}
The nonlocal source in (\ref{Bmode1}) is therefore only enhanced over the 
minuscule local contribution by a factor of $\ln(a)$. The net effect is no
large logarithms in 1-loop corrections to $u_B(\eta,k)$, just a slightly
slower approach to the constant late time limit of the tree order result,
\begin{equation}
u_B(\eta,k) \longrightarrow \Biggl\{ 1 + \frac{\lambda^2 H^2}{8 \pi^2}
\Bigl( \frac{k}{a H}\Bigr)^2 \ln(a) + O(\lambda^4) \Biggr\} u_0(\eta,k)
\; . \label{Bu}
\end{equation}

\subsubsection{1-Loop Corrected Exchange Potentials}

We define the exchange potential as the response to a point source
$J(\eta,\vec{x}) = K a \delta^3(\vec{x})$. These potentials are functions
of $\eta$ and $r \equiv \Vert \vec{x}\Vert$. The order $\lambda^0$ solutions
for $A$ and $B$ are the same \cite{Glavan:2019yfc},
\begin{equation}
\mathcal{D} P_0(\eta,r) = K a \delta^3(\vec{x}) \Longrightarrow
P_0(\eta,r) = \frac{K H}{4 \pi} \Bigl\{ \ln\Bigl(H r \!+\! \frac1{a}\Bigr) 
- \frac1{a H r}\Bigr\} \; . \label{P0new}
\end{equation}
Other derivatives of $P_0(\eta,r)$ are,
\begin{eqnarray}
\partial_0 P_0(\eta,r) &\!\!\! = \!\!\!& \frac{K H^2}{4\pi} \Bigl[ 
\frac1{H r} - \frac1{Hr \!+\! \frac1{a}}\Bigr] \;\; , \;\; \partial_0^2
P_0(\eta,r) = -\frac{K H^3}{4 \pi} \frac1{(Hr \!+\! \frac1{a})^2} \; , 
\qquad \\
\nabla^2 P_0(\eta,r) &\!\!\! = \!\!\!& \frac{K \delta^3(\vec{x})}{a} +
\frac{K H^3}{4 \pi} \Bigl[ \frac{2 a}{H r} - \frac{2 a}{Hr \!+\! \frac1{a}}
- \frac1{(Hr \!+\! \frac1{a})^2} \Bigr] \; . \qquad 
\end{eqnarray} 
Recall also that the late time limit of $H r \gg \frac1{a}$ is constant in 
time but not in space,
\begin{equation}
P_0(\eta,r) \longrightarrow \frac{K H}{4 \pi} \Bigl\{ \ln(H r) + 
\frac1{2 a^2 H^2 r^2} + \dots \Bigr\} \; . \label{lateP0}
\end{equation}

The exchange potential for $A$ takes the form,
\begin{equation}
P_A(\eta,r) = P_0(\eta,r) + P_{A1}(\eta,r) + O(\lambda^4) \; . 
\end{equation}
From equations (\ref{A1eqn}) and (\ref{P0new}) we see that $\mathcal{D} 
P_{A1}(\eta,r)$ is,
\begin{eqnarray}
\lefteqn{ a^4 H^2 m_A^2 P_0 - \frac{\lambda^2 H^2 \ln(\frac{ \mu a}{H}) K a 
\delta^3(\vec{x})}{16 \pi^2}  + \frac{\lambda^2 H^3 a^3 \partial_0 P_0}{16 \pi^2} 
- \frac{\lambda^2 \mathcal{D}}{64 \pi^2} \Bigl[ \frac{\ln(a) K \delta^3(\vec{x})}{
a^3}\Bigr] } \nonumber \\
& & \hspace{-0.5cm} - \frac{\lambda^2 K \mathcal{D}}{512 \pi^3} \!\! \int \!\! 
d^4\eta' \Biggl\{ \frac{\partial^4}{2 a^2 a'} \Biggl[ \theta(\Delta \eta \!-\! r) 
\Bigl( \ln[\mu^2 (\Delta \eta^2 \!-\! r^2)] \!-\! 1 \Bigr) \Biggr] 
+ \frac{2 H^2 \partial^2}{a} \Biggl[ \theta(\Delta \eta \!-\! r) \nonumber \\
& & \hspace{0cm} \times \Bigl(\ln\Bigl[\frac14 H^2 (\Delta \eta^2 \!-\! r^2) 
\Bigr] \!-\! 1\Bigr) \Biggr] - 4 a' H^4 \theta(\Delta \eta \!-\! r) \ln\Bigl[
\frac14 H^2 (\Delta \eta^2 \!-\! r^2)\Bigr] \Biggr\} . \qquad \label{PA1eqn}
\end{eqnarray}
It turns out that only the first two contributions to (\ref{PA1eqn}) make 
significant contributions to $P_{A1}(\eta,r)$ at late times. Of course the term 
proportional to $m_A^2 = \frac{3 \lambda^2 H^2}{32 \pi^2} = -m^2_{\Phi}$ makes 
the opposite contribution from the tachyonic mass of $\Phi$ that we worked out 
in expression (\ref{PhiP}),
\begin{equation}
\mathcal{D} \Delta P_{A1} = a^4 H^2 m_A^2 P_0 \Longrightarrow \Delta P_{A1}(\eta,r)
\longrightarrow -\frac{\lambda^2 H^2}{32 \pi^2} \ln(a) \!\times\! \frac{K H}{4 \pi} 
\ln(H r) \; . \label{PA1st}
\end{equation}
To work out the result from the second term in (\ref{PA1eqn}) we simply integrate
against the $\lambda^0$ retarded Green's function (\ref{Gretexp}),
\begin{eqnarray}
\lefteqn{ \mathcal{D} \Delta P_{A1} = - \frac{\lambda^2 H^2 \ln(\frac{ \mu a}{H}) 
K a \delta^3(\vec{x})}{16 \pi^2} } \nonumber \\
& & \hspace{-0.5cm} \Longrightarrow \Delta P_{A1}(\eta,r) = \int_{\eta_i}^{0} \!\! 
d\eta' \Bigl\{ \frac{\delta(\Delta \eta \!-\! r)}{a a' r} + H^2 \theta(\Delta \eta 
\!-\! r) \Bigr\} \frac{\lambda^2 K H^2 a' \ln(\frac{\mu a'}{H}) }{64 \pi^3} , 
\qquad \\
& & \hspace{2.3cm} = \frac{\lambda^2 K H^3}{64 \pi^3} \Bigl\{ \frac1{a H r} 
\ln\Bigl( \frac{\mu a(\eta \!-\! r)}{H}\Bigr) + \frac12 \ln^2\Bigl( 
\frac{\mu a(\eta \!-\! r)}{H}\Bigr) \Bigr\} , \qquad \\
& & \hspace{2.3cm} \longrightarrow \frac{\lambda^2 H^2}{32 \pi^2} \ln(Hr) \times
\frac{K H}{4 \pi} \ln(Hr) \; . \qquad \label{PA2nd}
\end{eqnarray}
The final step is facilitated by noting that $a(\eta - r) = 1/(Hr + \frac1{a})$.
Combining (\ref{PA1st}) and (\ref{PA2nd}) gives the leading late time correction
at 1-loop order,
\begin{equation}
P_{A1}(\eta,r) \longrightarrow \Bigl\{ -\frac{\lambda^2 H^2}{32 \pi^2} \ln(a) +
\frac{\lambda^2 H^2}{32 \pi^2} \ln(H r)\Bigr\} \times \frac{K H}{4\pi} \ln(Hr) 
\; . \label{APhi} 
\end{equation}

The $B$ exchange potential can be written as,
\begin{equation}
P_B(\eta,r) = P_0(\eta,r) + P_{B1}(\eta,r) + O(\lambda^4) \; . 
\end{equation}
Substituting the tree order solution (\ref{P0new}) in the generic 1-loop
equation (\ref{B1eqn}) implies that the 1-loop correction obeys,
\begin{eqnarray}
\lefteqn{\mathcal{D} P_{B1}(x) = \frac{\lambda^2 H^2 \ln(\frac{ \mu a}{H}) 
K a \delta^3(\vec{x})}{16 \pi^2} - \frac{\lambda^2 H^3 a^3 \partial_0 P_0}{16 \pi^2} 
- \frac{\lambda^2 \mathcal{D}}{32 \pi^2} \Bigl[ \frac{\ln(a) K \delta^3(\vec{x})}{
a^3}\Bigr] } \nonumber \\
& & \hspace{-0.5cm} - \frac{\lambda^2 K \mathcal{D}}{256 \pi^3} \! \int \!\!
d\eta' \Biggl\{ \frac{\partial^4}{2 a^2 a'} \Biggl[ \theta(\Delta \eta
\!-\! r) \Bigl( \ln[\mu^2 (\Delta \eta^2 \!-\! r^2)] \!-\! 1\Bigr) \Biggr] 
+ \frac{4 H^2 \partial^2}{a} \Bigl( \theta(\Delta \eta \!-\! r) 
\nonumber \\
& & \hspace{0.5cm} \times \ln\Bigl[\frac14 H^2 (\Delta \eta^2 \!-\! r^2)\Bigr] 
\Bigr) - 8 a' H^4 \theta(\Delta \eta \!-\! r) \Bigl(\ln\Bigl[\frac14 H^2 (\Delta 
\eta^2 \!-\! r^2)\Bigr] \!+\! 1\Bigr) \Biggr\} \nonumber \\
& & \hspace{-0.5cm} +\frac{\lambda^2 H^2 \partial^{\mu}}{128 \pi^3} \Biggl\{ a
\partial^4 \!\!\! \int \!\! d^4x' a' \theta(\Delta \eta \!-\! \Delta r) 
\Bigl[ \ln[\mu^2 (\Delta \eta^2 \!-\! \Delta x^2)] \!-\! 1\Bigr] \partial'_{\mu}
P_0(x') \Biggr\} \nonumber \\
& & \hspace{4cm} -\frac{\lambda^2 a^2 H^4 \partial^2}{64 \pi^3} \int \!\! 
d^4x' {a'}^2 \theta(\Delta \eta \!-\! \Delta r) \nabla^2 P_0(x')
\; . \qquad \label{PB1eqn}
\end{eqnarray}
Many of the contributions in (\ref{PB1eqn}) are similar to those of 
(\ref{PA1eqn}), and it turns out that only the first one induces large
logarithms at late times. Because the sign of this first term is opposite
to its cousin in (\ref{PA1eqn}) we need only reverse the sign of (\ref{PA2nd})
to obtain the leading late time contribution,
\begin{equation}
P_{B1}(\eta,r) \longrightarrow -\frac{\lambda^2 H^2}{32 \pi^2} \ln(H r) 
\times \frac{K H}{4\pi} \ln(Hr) \; . \label{BPhi} 
\end{equation}

\subsection{Expectation Values at 1 and 2 Loops}

Computing expectation values is {\it much} more difficult without a local 
field redefinition like (\ref{PsiofPhi}) which expresses the full field 
in terms of a free field. However, we struggle through most of the same 
computations for the two field model (\ref{LAB}) that we did for the 
single field model (\ref{LPhi}). We first evaluate primitive results in
dimensional regularization, then renormalize and take the unregulated 
limit.  

\subsubsection{Primitive Result for $\langle A(x)\rangle$}

\begin{figure}[H]
\centering
\includegraphics[width=5cm]{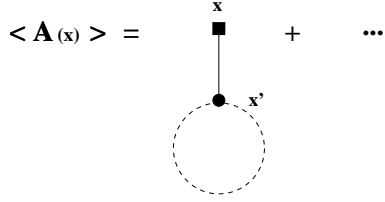}
\caption{\footnotesize Diagram which represents the 1-loop contribution to 
the expectation value of $A(x)$. Recall that $A$ lines are solid, whereas $B$
lines are dashed. Square vertices are fixed, whereas circular vertices are 
integrated.}
\label{figAB-A1}
\end{figure}

Figure~\ref{figAB-A1} shows the 1-loop contributions to the expectation
value of $A(x)$.
\noindent The initial expression of this diagram is,
\begin{equation}
A_{1} = -\frac{i}{2} \lambda \!\! \int \!\! d^Dx' \, i\Delta(x;x')
\sqrt{-g(x')} g^{\mu\nu}(x') \!\times\! \partial'_{\mu} \partial''_{\nu}
i\Delta(x';x'') \Bigr\vert_{x'' = x'} \; . \label{A1a}
\end{equation}
Employing relation (\ref{coincddprop}) to evaluate the doubly differentiated,
coincident propagator, and then interpreting the diagram in the 
Schwinger-Keldysh sense gives,
\begin{equation}
A_{1} = \frac{i}{2} \lambda (D\!-\!1) k H^2 \!\! \int \!\! d^Dx' 
\sqrt{-g(x')} \Bigl[ i\Delta_{++}(x;x') \!-\! i\Delta_{+-}(x;x')\Bigr] \; .
\label{A1b}
\end{equation}
Expression (\ref{A1b}) is ultraviolet finite so we can set $D=4$ and use
expressions (\ref{propD4}) and (\ref{rule2}) to conclude,
\begin{equation}
\Bigl\langle \Omega \Bigl\vert A(x) \Bigr\vert \Omega \Bigr\rangle =
\frac{\lambda H^2}{16 \pi^2} \Bigl[ \ln(a) - \frac13 + \frac1{3 a^3} \Bigr]
+ O(\lambda^3) \; . \label{AVEV}
\end{equation}

\subsubsection{Primitive Results for $\langle [A(x) - \langle A(x)\rangle]^2 
\rangle$}

\begin{figure}[H]
\centering
\includegraphics[width=11cm]{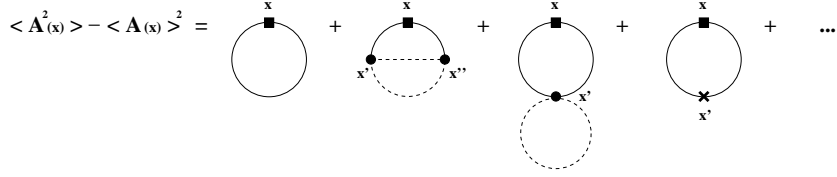}
\caption{\footnotesize Diagrams representing those 1-loop and 2-loop contributions
to the expectation value of $A^2(x)$ which are not already included in the square 
of the expectation value of $A(x)$. Recall that $A$ lines are solid, whereas $B$
lines are dashed. Square vertices are fixed, whereas circular vertices are 
integrated.}
\label{figAB-A2}
\end{figure}

The expectation value of $A^2(x)$ contains a disconnected part that is the 
square of (\ref{AVEV}),
\begin{equation}
\Bigl\langle \Omega \Bigl\vert A^2(x) \Bigr\vert \Omega \Bigr\rangle =
\Bigl\langle \Omega \Bigl\vert A(x) \Bigr\vert \Omega \Bigr\rangle^2 +  
\Bigl\langle \Omega \Bigl\vert \Bigl[A(x) - \langle \Omega \vert A(x) \vert
\Omega \rangle \Bigr]^2 \Bigr\vert \Omega \Bigr\rangle \; . \label{connected}
\end{equation}
Figure~\ref{figAB-A2} shows the 1-loop and 2-loop contributions to the connected
part.
\noindent Of course the 1-loop part is just the coincident propagator 
(\ref{coincprop}). Our initial expressions for the three 2-loop contributions 
are,
\begin{eqnarray}
A^2_{2a} &\!\!\! = \!\!\!& \frac{(-i \lambda)^2}{2} \!\! \int \!\! d^Dx' 
\sqrt{-g(x')} g^{\mu\nu}(x') \, i\Delta(x;x') \nonumber \\
& & \hspace{-0.5cm} \times \!\!\int \!\! d^Dx'' \sqrt{-g(x'')} g^{\rho\sigma}(x'')
\, i\Delta (x;x'') \, \partial'_{\mu} \partial''_{\rho} i\Delta(x';x'') 
\partial'_{\nu} \partial''_{\sigma} i\Delta(x';x'') \; , \qquad \label{A2a} \\
A^2_{2b} &\!\!\! = \!\!\!& -\frac{i \lambda^2}{4} \!\! \int \!\! d^Dx' 
\sqrt{-g(x')} g^{\mu\nu}(x') \Bigl[  i\Delta(x;x')\Bigr]^2 \partial'_{\mu} 
\partial''_{\nu} i\Delta(x';x'') \Bigl\vert_{x'' = x'} \; , \qquad 
\label{A2b} \\
A^2_{2c} &\!\!\! = \!\!\!& \int \!\! d^Dx' \sqrt{-g(x')} \Biggl\{ -i C_{A1} 
\square' i\Delta(x;x') \square' i\Delta(x;x') \nonumber \\
& & \hspace{4cm} - i C_{A2} \, R g^{\mu\nu}(x') \partial'_{\mu} i\Delta(x;x') 
\partial'_{\nu} i\Delta(x;x') \Biggr\} . \qquad \label{A2c}
\end{eqnarray}

We reduce expression (\ref{A2a}) with (\ref{ID2}) and an invocation of the
Schwinger-Keldysh formalism,
\begin{eqnarray}
A^2_{2a} &\!\!\! = \!\!\! & -\frac{\lambda^2}{8} \Bigl[ i\Delta(x;x)\Bigr]^2 
\nonumber \\
& & \hspace{-0.4cm} - \frac{i \lambda^2 (D \!-\! 1) k H^2}{4} \!\! \int \!\! 
d^Dx' \sqrt{-g(x')} \Biggl\{ \!\Bigl[ i\Delta(x;x') \Bigr]^2 \!\!\!-\! 
\Bigl[ i\Delta_{+-}(x;x') \Bigr]^2 \!\Biggr\} . \qquad \label{A2afinal}
\end{eqnarray}
The 4-point contribution (\ref{A2b}) follows from (\ref{coincddprop}) and
another application of the Schwinger-Keldysh formalsim,
\begin{equation}
A^2_{2b} = \frac{i \lambda^2 (D \!-\! 1) k H^2}{4} \!\! \int \!\! d^Dx' 
\sqrt{-g(x')} \Biggl\{ \!\Bigl[ i\Delta(x;x') \Bigr]^2 \!\!\!-\! \Bigl[ 
i\Delta_{+-}(x;x') \Bigr]^2 \!\Biggr\} . \label{A2bfinal}
\end{equation}
And the counterterm insertion (\ref{A2c}) follows from the propagator
equation and an application of (\ref{ID1}),
\begin{equation}
A^2_{2c} = - C_{A2} \, R \, i\Delta(x;x) \; . \label{A2cfinal}
\end{equation}

\subsubsection{Primitive Results for $\langle B^2(x) \rangle$}

\begin{figure}[H]
\centering
\includegraphics[width=11cm]{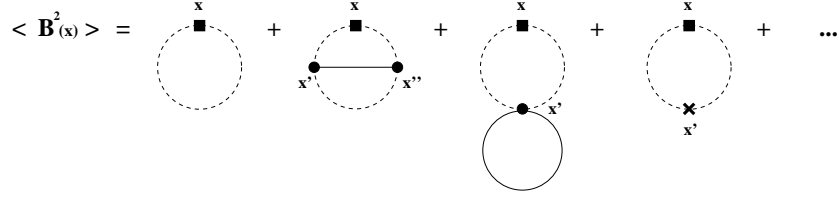}
\caption{\footnotesize Diagrams which represent the 1-loop and 2-loop contributions
to the expectation value of $B^2(x)$. Recall that $A$ lines are solid, whereas $B$
lines are dashed. Square vertices are fixed, whereas circular vertices are 
integrated.}
\label{figAB-B2}
\end{figure}

The expectation value of $B(x)$ vanishes to all orders by virtue of the
shift symmetry of (\ref{LAB}). Hence there is no disconnected part to the
expectation value of $B^2(x)$. The diagrams which contribute to it are
depicted in Figure~\ref{figAB-B2}
\noindent The associated analytic expressions are,
\begin{eqnarray}
B^2_{2a} &\!\!\! = \!\!\!& (-i \lambda)^2 \!\! \int \!\! d^Dx' \sqrt{-g(x')}
g^{\mu\nu}(x') \, \partial'_{\mu} i\Delta(x;x') \nonumber \\
& & \hspace{-0.5cm} \times \!\!\int \!\! d^Dx'' \sqrt{-g(x'')} g^{\rho\sigma}(x'')
\, \partial''_{\rho} i\Delta (x;x'') \, i\Delta(x';x'') \partial'_{\nu} 
\partial''_{\sigma} i\Delta(x';x'') \; , \qquad \label{B2a} \\
B^2_{2b} &\!\!\! = \!\!\!& -\frac{i \lambda^2}{4} \!\! \int \!\! d^Dx' 
\sqrt{-g(x')} g^{\mu\nu}(x') \partial'_{\mu} i\Delta(x;x') \partial'_{\nu} 
i\Delta(x;x') \, i\Delta(x';x') \; , \qquad \label{B2b} \\
B^2_{2c} &\!\!\! = \!\!\!& \int \!\! d^Dx' \sqrt{-g(x')} \Biggl\{ -i C_{B1} 
\square' i\Delta(x;x') \square' i\Delta(x;x') \nonumber \\
& & \hspace{4cm} - i C_{B2} \, R g^{\mu\nu}(x') \partial'_{\mu} i\Delta(x;x') 
\partial'_{\nu} i\Delta(x;x') \Biggr\} . \qquad \label{B2c}
\end{eqnarray}

Familiar partial integration procedures reduce (\ref{B2a}-\ref{B2c}) to the
form,
\begin{eqnarray}
B^2_{2a} &\!\!\! = \!\!\! & \frac{3 \lambda^2}{4} \Bigl[ i\Delta(x;x)\Bigr]^2 
\nonumber \\
& & \hspace{-0.4cm} - \frac{i \lambda^2 (D \!-\! 1) k H^2}{2} \!\! \int \!\! 
d^Dx' \sqrt{-g(x')} \Biggl\{ \!\Bigl[ i\Delta(x;x') \Bigr]^2 \!\!\!-\! 
\Bigl[ i\Delta_{+-}(x;x') \Bigr]^2 \!\Biggr\} , \qquad \label{B2afinal} \\
B^2_{2b} &\!\!\! = \!\!\!& -\frac{\lambda^2}{4} \Bigl[ i\Delta(x;x)\Bigr]^2 
\nonumber \\
& & \hspace{-0.4cm} + \frac{i \lambda^2 (D \!-\! 1) k H^2}{4} \!\! \int \!\! 
d^Dx' \sqrt{-g(x')} \Biggl\{ \!\Bigl[ i\Delta(x;x') \Bigr]^2 \!\!\!-\! 
\Bigl[ i\Delta_{+-}(x;x') \Bigr]^2 \!\Biggr\} , \qquad \label{B2abfinal} \\
B^2_{2c} &\!\!\! = \!\!\!& - C_{B2} \, R \, i\Delta(x;x) \; . \label{B2cfinal}
\end{eqnarray}
The coincident propagator was given in (\ref{coincprop}), so we need only
employ the Schwinger-Keldysh formalism to show that,
\begin{eqnarray}
\lefteqn{ - \frac{i \lambda^2 (D \!-\! 1) k H^2}{4} \!\! \int \!\! d^Dx' 
\sqrt{-g(x')} \Biggl\{ \!\Bigl[ i\Delta(x;x') \Bigr]^2 \!\!\!-\! \Bigl[ 
i\Delta_{+-}(x;x') \Bigr]^2 \!\Biggr\} } \nonumber \\
& & \hspace{-0.5cm} = \frac{\lambda^2 H^{2D-4}}{(4 \pi)^D} \frac{\Gamma(D)}{
\Gamma(\frac{D}2)} \Biggl\{ \frac{\Gamma(\frac{D}2 \!-\! 1)}{2 (D \!-\! 3)
(D \!-\! 4)} \!-\! \frac{\ln^2(a)}{3} \!-\! \frac23 \ln(a) \!+\! \frac{79}{54}
+ O(a^{-1}) \Biggr\} . \qquad \label{SKint}
\end{eqnarray}
Note that we have simplified ultraviolet finite contributions to (\ref{SKint})
by taking $D=4$.

\subsubsection{Renormalized Results}

The expectation value of $A(x)$ is ultraviolet finite and requires no
renormalization. The square of $A(x)$ is a composite operator and renormalizing 
it at 1-loop and 2-loop orders requires three counterterms,
\begin{equation}
\delta A^2 = K_{A1} R + K_{A2} R A^2 + K_{A3} R^2 + \dots \label{deltaA2}
\end{equation}
Comparison with expressions (\ref{A2afinal}-\ref{A2cfinal}) reveals that the
three constants are, 
\begin{eqnarray}
K_{A1} &\!\!\! = \!\!\!& \frac{\mu^{D-4}}{(4 \pi)^{\frac{D}2}} 
\frac{\Gamma(D\!-\!1)}{\Gamma(\frac{D}2)} \frac{\pi {\rm cot}(\frac{D \pi}{2})}{
D (D \!-\! 1)} \; , \label{KA1} \\
K_{A2} &\!\!\! = \!\!\!& 0 \; , \label{KA2} \\
K_{A3} &\!\!\! = \!\!\!& -\frac{\lambda^2 \mu^{2D-8}}{8 (4\pi)^{D}}
\frac{\Gamma^2(D\!-\!1)}{\Gamma^2(\frac{D}2)} \frac{\pi^2 {\rm cot}^2(
\frac{D \pi}{2})}{D^2 (D\!-\!1)^2} \; . \label{KA3}
\end{eqnarray}
Putting everything together gives the final renormalized result,
\begin{eqnarray}
\lefteqn{\Bigl\langle \Omega \Bigl\vert A^2(x)\Bigr\vert \Omega 
\Bigr\rangle_{\rm ren} = \frac{H^2}{4 \pi^2} \ln\Bigl( \frac{\mu a}{H}
\Bigr)- \frac{\lambda^2 H^4}{128 \pi^4} \ln^2\Bigl( \frac{\mu a}{H}\Bigr) }
\nonumber \\
& & \hspace{4cm} \frac{\lambda^2 H^4}{256 \pi^4} \Biggl[\ln(a) \!-\! \frac13
\!+\! \frac1{3 a^3} + O(\lambda^2)\Biggr]^2 + O(\lambda^4) \; . \qquad 
\label{A2ren}
\end{eqnarray}

The square of $B(x)$ is also a composite operator and requires similar
counterterms,
\begin{equation}
\delta B^2 = K_{B1} R + K_{B2} R B^2 + K_{B3} R^2 + \dots \label{deltaB2}
\end{equation}
Comparison with expressions (\ref{B2afinal}-\ref{B2cfinal}) determines the
constants to be,
\begin{eqnarray}
K_{B1} &\!\!\! = \!\!\!& \frac{\mu^{D-4}}{(4 \pi)^{\frac{D}2}} 
\frac{\Gamma(D\!-\!1)}{\Gamma(\frac{D}2)} \frac{\pi {\rm cot}(\frac{D \pi}{2})}{
D (D \!-\! 1)} \; , \label{KB1} \\
K_{B2} &\!\!\! = \!\!\!& \frac{5 \lambda^2 \mu^{D-4}}{4 (4 \pi)^{\frac{D}2}}
\frac{\Gamma(D\!-\!1)}{\Gamma(\frac{D}2)} \frac{\pi {\rm cot}(\frac{D\pi}{2})}{
D (D\!-\!1)} - \frac{\lambda^2 \mu^{D-4}}{32 \pi^{\frac{D}2}} 
\frac{\Gamma(\frac{D}2 \!-\!1)}{2 (D\!-\!3) (D\!-\!4)} \Bigl( \frac{D\!-\!2}{D\!-\!1}
\Bigr) , \qquad \label{KB2} \\
K_{B3} &\!\!\! = \!\!\!& \frac{\lambda^2 \mu^{2D-8}}{2 (4\pi)^{D}}
\frac{\Gamma^2(D\!-\!1)}{\Gamma^2(\frac{D}2)} \frac{\pi^2 {\rm cot}^2(
\frac{D \pi}{2})}{D^2 (D\!-\!1)^2} \nonumber \\
& & \hspace{0.5cm} - \frac{\lambda^2 \mu^{2D-8}}{(4\pi)^{D}}
\frac{\Gamma(D\!-\!1)}{\Gamma(\frac{D}2)} \frac{\Gamma(\frac{D}2 \!-\! 1)}{2
(D\!-\!3) (D\!-\!4)} \frac{1}{D^2 (D \!-\!1)} - \frac{\lambda^2}{128 \pi^4}
\frac{79}{2592} \; . \qquad \label{KB3}
\end{eqnarray}
Substituting these values in (\ref{deltaB2}), adding that to the sum of
(\ref{B2afinal}-\ref{B2cfinal}), and taking the unregulated limit gives,
\begin{eqnarray}
\lefteqn{\Bigl\langle \Omega \Bigl\vert B^2(x) \Bigr\vert \Omega 
\Bigr\rangle_{\rm ren} = \frac{H^2}{4 \pi^2} \ln\Bigl( \frac{\mu a}{H}
\Bigr) + \frac{\lambda^2 H^4}{128 \pi^4} \Biggl\{ 3 \ln^2(a) \!-\! 
2 \ln(a) } \nonumber \\
& & \hspace{3cm} + 8 \ln\Bigl( \frac{\mu}{H} \Bigr) \ln(a) \!+\! 
4 \ln^2\Bigl(\frac{\mu}{H}\Bigr) \!-\! 3 \ln\Bigl(\frac{\mu}{H}\Bigr)
\Biggr\} + O(\lambda^4) \; . \qquad \label{B2ren}
\end{eqnarray}

\section{Describing the Large Logarithms}

The purpose of this section is to explain the various large logarithms
derived in the previous two sections. We begin by summarizing them. We
then show how to infer stochastic effects based on effective potentials
for $\Phi(x)$ and $A(x)$. The remaining large logarithms can be explained
using a variant of the renormalization group which is based on a special
class of counterterms that can be viewed as curvature-dependent 
renormalizations of parameters in the bare theories. The section closes
with a color-coded summary of the various logarithms, in which stochastic 
effects are red and renormalization group effects are green.

\subsection{Summary}

The large logarithms of the single scalar model (\ref{LPhi}) reside
in expressions (\ref{Phiu}), (\ref{PhiP}), (\ref{PhiVEV}) and
(\ref{Phi2ren}). Table~\ref{SingleLogs} summarizes these results.
\begin{table}[H]
\setlength{\tabcolsep}{8pt}
\def\arraystretch{1.5}
\centering
\begin{tabular}{|@{\hskip 1mm }c@{\hskip 1mm }||c|}
\hline
Quantity & Leading Logarithms \\
\hline\hline
$u_{\Phi}(\eta,k)$ & $\Bigl\{1 + \frac{\lambda^2 H^2}{32 \pi^2} \ln(a)
+ O(\lambda^4)\Bigr\} \times \frac{H}{\sqrt{2 k^3}}$ \\
\hline
$P_{\Phi}(\eta,r)$ & $\Bigl\{1 + \frac{\lambda^2 H^2}{32 \pi^2} \ln(a) 
+ O(\lambda^4)\Bigr\} \times \frac{KH}{4\pi} \ln(Hr)$ \\
\hline
$\langle \Omega \vert \Phi(x)\vert \Omega \rangle$ & $-\Bigl\{1 + 
\frac{15 \lambda^2 H^2}{64 \pi^2} \ln(a) + O(\lambda^4)\Bigr\} \times 
\frac{\lambda H^2}{16 \pi^2} \ln(a)$ \\
\hline
$\langle \Omega \vert \Phi^2(x)\vert \Omega \rangle_{\rm ren}$ & 
$\Bigl\{1 + \frac{15 \lambda^2 H^2}{64 \pi^2} \ln(a) + O(\lambda^4)\Bigr\} 
\times \frac{H^2}{4 \pi^2} \ln(a)$ \\
\hline
\end{tabular}
\caption{\footnotesize Leading logarithms in the single scalar model
(\ref{LPhi}).}
\label{SingleLogs}
\end{table}

The large logarithms of the two scalar model (\ref{LAB}) reside
in expressions (\ref{Au}), (\ref{Bu}), (\ref{APhi}), (\ref{BPhi}),
(\ref{AVEV}), (\ref{A2ren}) and (\ref{B2ren}). Of course the expectation
value of $B(x)$ vanishes to all orders. Table~\ref{DoubleLogs} 
summarizes these results.
\begin{table}[H]
\setlength{\tabcolsep}{8pt}
\def\arraystretch{1.5}
\centering
\begin{tabular}{|@{\hskip 1mm }c@{\hskip 1mm }||c|}
\hline
Quantity & Leading Logarithms \\
\hline\hline
$u_{A}(\eta,k)$ & $\Bigl\{1 - \frac{\lambda^2 H^2}{32 \pi^2} \ln(a) 
+ O(\lambda^4)\Bigr\} \times \frac{H}{\sqrt{2 k^3}}$ \\
\hline
$u_{B}(\eta,k)$ & $\Bigl\{1 + 0 + O(\lambda^4)\Bigr\} \times 
\frac{H}{\sqrt{2 k^3}}$ \\
\hline
$P_{A}(\eta,r)$ & $\Bigl\{1 - \frac{\lambda^2 H^2}{32 \pi^2} \ln(a)
+ \frac{\lambda^2 H^2}{32 \pi^2} \ln(Hr) + O(\lambda^4)\Bigr\} \times 
\frac{KH}{4\pi} \ln(Hr)$ \\
\hline
$P_{B}(\eta,r)$ & $\Bigl\{1 - \frac{\lambda^2 H^2}{32 \pi^2} \ln(Hr) 
+ O(\lambda^4)\Bigr\} \times \frac{KH}{4\pi} \ln(Hr)$ \\
\hline
$\langle \Omega \vert A(x)\vert \Omega \rangle$ & $\Bigl\{1 + 
O(\lambda^2)\Bigr\} \times \frac{\lambda H^2}{16 \pi^2} \ln(a)$ \\
\hline
$\langle \Omega \vert A^2(x)\vert \Omega \rangle_{\rm ren}$ & 
$\Bigl\{1 - \frac{\lambda^2 H^2}{64 \pi^2} \ln(a) + O(\lambda^4)\Bigr\} 
\times \frac{H^2}{4 \pi^2} \ln(a)$ \\
\hline
$\langle \Omega \vert B(x)\vert \Omega \rangle$ & $0$ \\
\hline
$\langle \Omega \vert B^2(x)\vert \Omega \rangle_{\rm ren}$ & 
$\Bigl\{1 + \frac{3 \lambda^2 H^2}{32 \pi^2} \ln(a) + O(\lambda^4)
\Bigr\} \times \frac{H^2}{4 \pi^2} \ln(a)$ \\
\hline
\end{tabular}
\caption{\footnotesize Leading logarithms in the two scalar model
(\ref{LAB}).}
\label{DoubleLogs}
\end{table}

\subsection{1-Loop Effective Potentials}

Stochastic effects in both models can understood using the effective
potential. One derives this by setting the undifferentiated fields 
equal to a constant and then integrating the differentiated fields 
out of the field equations.

\subsubsection{Single Field Model}

The key to evaluating $V_{\rm eff}(\Phi)$ is that the $\Phi$ 
propagator in the presence of a constant field configuration $\Phi(x) = 
\Phi_0$ is a field strength renormalization of the free propagator,
\begin{equation}
\Bigl\langle \Omega \Bigl\vert T\Bigl[ \Phi(x) \Phi(x')\Bigr] \Bigr\vert 
\Omega \Bigr\rangle_{\Phi_0} = \frac{i\Delta(x;x')}{(1 \!+\! \frac12 \lambda
\Phi_0)^2} \; . \label{Phifullprop}
\end{equation}
The 1-loop effective potential follows from taking the expectation value
of the action's first variation (\ref{varPhi1}) in the presence of constant
$\Phi(x) = \Phi_0$,
\begin{eqnarray}
-V'_{\rm eff}(\Phi_0) a^{D} &\!\!\! = \!\!\!& \Bigl(1 \!+\! \frac12 
\lambda \Phi_0 \Bigr) \partial^{\mu} \Bigl[ \frac14 \lambda a^{D-2} 
\partial_{\mu} \Bigl\langle \Omega \Bigl\vert \Phi^2(x)
\Bigr\vert \Omega \Bigr\rangle_{\Phi_0} \Bigr] \; , \qquad \\
&\!\!\! = \!\!\!& - \frac{\frac12 \lambda (D \!-\! 1) k H^2 a^D}{1
\!+\! \frac12 \lambda \Phi_0} \; . \qquad \label{VPhiprime} 
\end{eqnarray}
Expression (\ref{VPhiprime}) is ultraviolet finite and corresponds to a
1-loop effective potential of,
\begin{equation}
V_{\rm eff}(\Phi) = \frac{3 H^4}{8 \pi^2} \ln\Bigl\vert 1 + \frac12 
\lambda \Phi\Bigr\vert \; . \label{VPhi}
\end{equation}

The effective potential (\ref{VPhi}) explains the tachyonic mass 
(\ref{MPhi}) we found after the lengthy computation of the 1-loop self-mass, 
\begin{equation}
m^2_{\Phi} \equiv \frac{\partial^2 V_{\rm eff}(\Phi)}{\partial \Phi^2} 
\Bigl\vert_{\Phi = 0} = -\frac{3 \lambda^2 H^4}{32 \pi^2} \; .
\label{m2Phi}
\end{equation}
Recall expression (\ref{Phiu}) for the late time limit of the massive 
mode function,
\begin{equation}
u_{\Phi}(\eta,k) \longrightarrow \frac{\Gamma(\nu)}{\sqrt{4 \pi H a^3}}
\Bigl( \frac{2 a H}{k}\Bigr)^{\nu} \qquad , \qquad \nu = \sqrt{\frac94
- \frac{m^2_{\Phi}}{H^2} } = \frac32 - \frac{m^2_{\Phi}}{3 H^2} + \dots
\label{uPhilate}
\end{equation}
Substituting (\ref{m2Phi}) into (\ref{uPhilate}) and expanding for small
$\lambda$ gives quantitative agreement with the entry for the mode function
in Table~\ref{SingleLogs},
\begin{equation}
u_{\Phi}(\eta,k) \longrightarrow \Biggl\{ 1 + \frac{\lambda^2 H^2}{32 \pi^2}
\Bigl[ \ln\Bigl(\frac{2 a H}{k}\Bigr) + \psi\Bigl(\frac32\Bigr)\Bigr] + 
O(\lambda^4) \Biggr\} \frac{H}{\sqrt{2 k^3}} \; . \label{finaluPhi}
\end{equation}
From expression (\ref{PhiP}) we see that the stochastically generated mass
(\ref{m2Phi}) also explains the entry for the exchange potential.

The effective potential also explains the tendency for the expectation 
value of $\Phi(x)$ to become more and more negative as per expression 
(\ref{PhiVEV}). To see this we write the homogeneous evolution equation 
(in co-moving time) that follows from adding the variation of the effective 
potential to the classical variation (\ref{varPhi1}),
\begin{equation}
-\Bigl(1 \!+\! \frac{\lambda}{2} \Phi\Bigr) \frac{d}{dt} \Biggl[\Bigl(1 
\!+\! \frac{\lambda}{2} \Phi\Bigr) a^3 \dot{\Phi}\Biggr] -
V'_{\rm eff}(\Phi) a^3 = 0 \; . \label{Phieffeqn}
\end{equation}
Because the evolution of $\Phi$ is much slower than that of the scale 
factor $a = e^{H t}$, the largest contribution to the first term of
(\ref{Phieffeqn}) is from the external derivative acting on the factor of 
$a^3$. At this point the equation can be integrated,
\begin{equation}
3 H \Bigl(1 \!+\! \frac{\lambda}{2} \Phi\Bigr)^2 \dot{\Phi} \simeq 
-\frac{3 \lambda H^4}{16 \pi^2} \frac1{1 \!+\! \frac12 \lambda \Phi} 
\Longrightarrow \frac1{2\lambda} \Bigl[ \Bigl(1 \!+\! \frac{\lambda}{2}
\Phi\Bigr)^4 - 1\Bigr] \simeq -\frac{\lambda H^2}{16 \pi^2} \ln(a) \; . 
\label{Phievolution}
\end{equation}
Inverting to solve for $\Phi$ gives,
\begin{equation}
\Phi = \frac{2}{\lambda} \Biggl\{ \Bigl[1 - \frac{\lambda^2 H^2}{8 \pi^2}
\ln(a)\Bigr]^{\frac14} - 1 \Biggr\} = -\frac{\lambda H^2}{16 \pi^2} \ln(a) 
\Biggl\{1 + \frac{3 \lambda^2 H^2}{64 \pi^2} \ln(a) + O(\lambda^4) \Biggr\} .
\label{classPhi}
\end{equation}

The fact that the order $\lambda^3$ contribution (\ref{classPhi}) 
disagrees with Table~\ref{SingleLogs} is due to not having included 
fluctuations around the homogeneous solution driven by the stochastically
truncated free field,
\begin{equation}
\varphi_0(t,\vec{x}) \equiv \int \!\! \frac{d^3k}{(2\pi)^3} \theta\Bigl(a H \!-\!
k\Bigr) \frac{\theta(k \!-\! H) H}{\sqrt{2 k^3}} \Bigl\{ \alpha_{\vec{k}} 
e^{i \vec{k} \cdot\vec{x}} + \alpha^{\dagger}_{\vec{k}} e^{-i \vec{k} \cdot 
\vec{x}} \Bigr\} \; , \label{stochfree}
\end{equation}
where $\alpha^{\dagger}_{\vec{k}}$ and $\alpha_{\vec{k}}$ are canonically
normalized creation and annihilation operators,
\begin{equation}
\Bigl[ \alpha_{\vec{k}} , \alpha^{\dagger}_{\vec{p}} \Bigr] = (2\pi)^3
\delta^3(\vec{k} \!-\! \vec{p}) \; .
\end{equation}
If we use the symbol $\varphi(t,\vec{x})$ to distinguish the full ultraviolet 
finite stochastic field from $\Phi$, then the Langevin equation associated 
with (\ref{Phievolution}) is,
\begin{equation}
\dot{\varphi} = \dot{\varphi}_0 - \frac{\lambda H^3}{16 \pi^2} \frac1{(1 \!+\! 
\frac12 \lambda \Psi)^3} \; . \label{Langevin}
\end{equation}
It is simple to generate a perturbative solution which includes stochastic
fluctuations around (\ref{classPhi}),
\begin{eqnarray}
\lefteqn{\varphi = \varphi_0 -\frac{\lambda H^2}{16 \pi^2} \ln(a) + 
\frac{3 \lambda^2 H^3}{32 \pi^2} \int_{0}^{t} \!\! dt' \varphi_0 } 
\nonumber \\
& & \hspace{4cm} - \frac{3 \lambda^3 H^4}{1024 \pi^4} \ln^2(a) - 
\frac{3 \lambda^3 H^3}{32 \pi^2} \int_{0}^{t} \!\! dt' \varphi_0^2 
+ O(\lambda^4) \; . \qquad \label{stochexp}
\end{eqnarray}
The expectation value of (\ref{stochexp}) reproduces the entry for $\langle
\Omega \vert \Phi(x) \vert \Omega \rangle$ in Table~\ref{SingleLogs},
\begin{eqnarray}
\Bigl\langle \Omega \Bigl\vert \varphi(t,\vec{x}) \Bigr\vert \Omega 
\Bigr\rangle &\!\!\! = \!\!\!& 0 - \frac{\lambda H^2}{16 \pi^2} \ln(a) + 0 
\nonumber \\
& & \hspace{0.4cm} - \frac{3 \lambda^3 H^4}{1024 \pi^4} \ln^2(a) - 
\frac{3 \lambda^3 H^3}{32 \pi^2} \int_{0}^{t} \!\! dt' \frac{H^2}{4 \pi^2} 
\ln(a') + O(\lambda^5) \; , \qquad \\
&\!\!\! = \!\!\!& -\frac{\lambda H^2}{16 \pi^2} \ln(a) \Biggl\{1 + 
\frac{15 \lambda^2 H^2}{64 \pi^4} \ln^2(a) + O(\lambda^4) \Biggr\} . 
\qquad \label{stochVEV}
\end{eqnarray}
Note that stochastic fluctuations cause the expectation value of the field 
to roll down its potential more rapidly than the result (\ref{classPhi}) 
because a downward fluctuation is more probable than an upward one.

\subsubsection{Two Field Model}

The same techniques can be applied to the two scalar model (\ref{LAB}). The 
exact shift symmetry of $B$ precludes there being any effective potential 
for the field $B$, but $A$ has one. We can compute it by noting that the 
expectation value of $B$ in the presence of constant $A(x) = A_0$ is a field 
strength renormalization,
\begin{equation}
\Bigl\langle \Omega \Bigl\vert T\Bigl[ B(x) B(x')\Bigr] \Bigr\vert \Omega
\Bigr\rangle_{A_0} = \frac{i\Delta(x;x')}{(1 \!+\! \frac12 \lambda A_0)^2} 
\; . \label{Bfullprop}
\end{equation}
The 1-loop effective potential follows from taking the expectation value
of the action's first A variation (\ref{Avar1}),
\begin{eqnarray}
-V'_{\rm eff}(A_0) a^D &\!\!\! = \!\!\!& -\frac12 \lambda \Bigl(1 \!+\! 
\frac12 \lambda A_0\Bigr) a^{D-2} \Bigl\langle \Omega \Bigl\vert \partial^{\mu}
B(x) \partial_{\mu} B(x) \Bigr\vert \Omega \Bigr\rangle_{A_0} \; , \qquad \\
&\!\!\! = \!\!\!& + \frac{\frac12 \lambda (D\!-\! 1) k H^2 a^D}{1 \!+\! \frac12
\lambda A_0} \; . \label{VAprime}
\end{eqnarray}
Taking the unregulated limit and integrating gives the 1-loop effective 
potential,
\begin{equation}
V_{\rm eff}(A) = -\frac{3 H^4}{8 \pi^2} \ln\Bigl\vert 1 + \frac12 \lambda A
\Bigr\vert \; . \label{VA}
\end{equation}

The $A$ effective potential (\ref{VA}) explains the positive mass-squared 
we found in expression (\ref{MAren}) after a lengthy computation,
\begin{equation}
m^2_{A} = \frac{\partial^2 V_{\rm eff}(A)}{\partial A^2} \Bigl\vert_{A = 0}
= \frac{3 \lambda^2 H^4}{32 \pi^2} \; . 
\end{equation}
This is opposite of the tachyonic mass for $\Phi$ (that is, $m^2_{A} = 
-m^2_{\Phi}$), which explains the factors of $-\frac{\lambda^2 H^2}{32 \pi^2} 
\ln(a)$ in the entries for $u_A(\eta,k)$ and $P_A(\eta,r)$ in 
Table~\ref{DoubleLogs}. The effective potential for $A$ also explains the
tendency for $\langle \Omega \vert A(x)\vert \Omega \rangle$ to grow without
bound. Specializing the $A$ field equation (\ref{Avar1}) to homogeneous 
evolution in co-moving coordinates, adding the effective potential, and 
neglecting derivatives of $A$ with respect to derivatives of the scale factor 
$a$ gives,
\begin{equation}
-\frac{d}{dt} \Bigl(  a^3 \dot{A}\Bigr) - V_{\rm eff}'(A) a^3 = 0 \qquad
\Longrightarrow \qquad 3 H \dot{A} \simeq \frac{3 \lambda H^4}{16 \pi^2}
\frac1{1 \!+\! \frac12 \lambda A} \; . \label{Aevolution}
\end{equation}
Equation (\ref{Aevolution}) can be solved exactly,
\begin{equation}
A \simeq \frac{2}{\lambda} \Biggl[ \sqrt{1 + \frac{\lambda^2 H^2}{16 \pi^2}
\ln(a)} - 1\Biggr] = \frac{\lambda H^2}{16 \pi^2} \ln(a) \Biggl\{1 -
\frac{\lambda^2 H^2}{64 \pi^2} \ln(a) + O(\lambda^4) \Biggr\} . \label{Asoln}
\end{equation}

Gaining quantitative agreement with $\langle \Omega \vert A(x)\vert \Omega
\rangle$ requires the inclusion of stochastic jitter from the truncated free
field $\mathcal{A}_0(t,\vec{x})$,
\begin{equation}
\mathcal{A}_0(t,\vec{x}) \equiv \int \!\! \frac{d^3k}{(2\pi)^3} 
\theta\Bigl(a H \!-\! k\Bigr) \frac{\theta(k \!-\! H) H}{\sqrt{2 k^3}} 
\Bigl\{ \alpha_{\vec{k}} e^{i \vec{k} \cdot\vec{x}} + \alpha^{\dagger}_{\vec{k}} 
e^{-i \vec{k} \cdot \vec{x}} \Bigr\} \; . \label{stochfreeA}
\end{equation}
The Langevin equation associated with (\ref{Aevolution}) is,
\begin{equation}
\dot{\mathcal{A}} = \dot{\mathcal{A}}_0 + \frac{\lambda H^3}{16 \pi^2} 
\frac1{1 \!+\! \frac12 \lambda \mathcal{A}} \; . \label{stochAeqn}
\end{equation}
Iteration of (\ref{stochAeqn}) generates a solution which includes the 
$\comp$-number solution (\ref{Asoln}) plus stochastic jitter involving
$\mathcal{A}_0$,
\begin{eqnarray}
\lefteqn{\mathcal{A} = \mathcal{A}_0 + \frac{\lambda H^2}{16 \pi^2} \ln(a)
- \frac{\lambda^2 H^3}{32 \pi^2} \int_{0}^{t} \!\! dt' \mathcal{A}_0 }
\nonumber \\
& & \hspace{4cm} - \frac{\lambda^3 H^4}{1024 \pi^4} \ln^2(a) + 
\frac{\lambda^3 H^3}{64 \pi^2} \int_{0}^{t} \!\! dt' \mathcal{A}^2_0 +
O(\lambda^4) \; . \qquad \label{stochAexp}
\end{eqnarray}
The expectation value of (\ref{stochAexp}) agrees exactly with the entry
for $\langle \Omega \vert A(x) \vert \Omega \rangle$ in Table~\ref{DoubleLogs},
\begin{eqnarray}
\Bigl\langle \Omega \Bigl\vert \mathcal{A}(t,\vec{x}) \Bigr\vert \Omega
\Bigr\rangle &\!\!\! = \!\!\!& 0 + \frac{\lambda H^2}{16 \pi^2} \ln(a) + 0
\nonumber \\
& & \hspace{0.5cm} - \frac{\lambda^3 H^4}{1024 \pi^4} \ln^2(a) + 
\frac{\lambda^3 H^3}{64 \pi^2} \int_{0}^{t} \!\! dt' \frac{H^2}{4 \pi^2}
\ln(a') + O(\lambda^5) \; , \qquad \\
&\!\!\! = \!\!\!& \frac{\lambda H^2}{16 \pi^2} \ln(a) \Biggl\{1 + 
\frac{\lambda^2 H^2}{64 \pi^2} \ln(a) + O(\lambda^4) \Biggr\} .
\label{stochAVEV}
\end{eqnarray}
Note again that stochastic jitter again increases the rate at which the field 
rolls down its potential.

The expectation value of $\mathcal{A}^2(t,\vec{x})$ is also straightforward,
\begin{eqnarray}
\Bigl\langle \Omega \Bigl\vert \mathcal{A}^2(t,\vec{x}) \Bigr\vert \Omega
\Bigr\rangle &\!\!\! = \!\!\!& \Biggl\langle \Omega \Biggl\vert \mathcal{A}_0^2
+ \frac{\lambda H^2}{8 \pi^2} \ln(a) \mathcal{A}_0 \nonumber \\
& & \hspace{1cm} + \frac{\lambda^2 H^4}{256 \pi^4} \ln^2(a) - 
\frac{\lambda^2 H^3}{16 \pi^2} \mathcal{A}_0 \int_{0}^{t} \!\! dt' \mathcal{A}_0
+ O(\lambda^3) \Biggr\rangle , \qquad \\
&\!\!\! = \!\!\!& \frac{H^2}{4 \pi^2} \ln(a) \Biggl\{ 1 - 
\frac{\lambda^2 H^2}{64 \pi^2} \ln(a) + O(\lambda^4) \Biggr\} . 
\label{stochA2VEV}
\end{eqnarray}
Expression (\ref{stochA2VEV}) is in perfect agreement with the entry for
$\langle \Omega \vert A^2(x) \vert \Omega \rangle$ in Table~\ref{DoubleLogs}. 
This means that the leading logarithms of the field $A(x)$ and its square are 
purely stochastic --- at least to this order.

\subsection{Curvature-Dependent Renormalizations}

Let us start with the renormalized expectation of $\Phi^2(x)$. Recall from
relation (\ref{deltaPhi2}) that the renormalized composite operator is,
\begin{equation}
\Phi^2_{\rm ren} \equiv \Phi^2 + K_{\Phi 1} R + K_{\Phi 2} R \Phi^2 + 
K_{\Phi 3} R^2 + O(\lambda^4) \; . \label{renPhi2}
\end{equation}
Some of the 1-loop and 2-loop counterterms in expression (\ref{renPhi2})
have no flat space analogs, but the $K_{\Phi 2} R \Phi^2$ counterterm can be 
regarded as part of a curvature-dependent field strength renormalization, 
$\Phi^2 = \sqrt{Z_{\Phi^2}} \times \Phi^2_{\rm ren}$ with,
\begin{equation}
Z_{\Phi^2} = 1 - 2 K_{\Phi 2} \!\times\! R + O(\lambda^4) \; . \label{ZPhi2}
\end{equation}
The associated $\gamma$ function is,
\begin{equation}
\gamma_{\Phi^2} \equiv \frac{\partial \ln(Z_{\Phi^2})}{\partial \ln(\mu^2)} 
= -\frac{15 \lambda^2 H^2}{32 \pi^2} + O(\lambda^4) \; . \label{gammaPhi2}
\end{equation}
Expression (\ref{Phi2ren}) shows that the renormalized expectation value of 
$\Phi^2(x)$ actually depends on the product $\mu a/H$. Hence we can replace 
the $\mu \frac{\partial}{\partial \mu}$ term in the Callan-Symanzik equation
with $a \frac{\partial}{\partial a}$,
\begin{equation}
\Bigl[ a \frac{\partial}{\partial a} + \beta \frac{\partial}{\partial \lambda}
+ \gamma_{\Phi^2} \Bigr] \Bigl\langle \Omega \Bigl\vert \Phi^2(x) \Bigr\vert
\Omega \Bigr\rangle_{\rm ren} = 0 \; . \label{Phi2CZeqn}
\end{equation}
The $\beta$ function for this model is of order $\lambda^3$, so we see that
equation (\ref{Phi2CZeqn}) perfectly explains the order $\lambda^2$ (2-loop)
contribution to $\langle \Omega \vert \Phi^2(x) \vert \Omega \rangle_{\rm ren}$
\begin{equation}
a \frac{\partial}{\partial a} \Bigl\{ \frac{15 \lambda^2 H^4}{256 \pi^4} 
\ln^2(a)\Bigr\} - \frac{15 \lambda^2 H^2}{32 \pi^2} \times \frac{H^2}{4 \pi^2} 
\ln(a) = 0 \label{explainCZeqn}
\end{equation}
Note that the order $\lambda^0$ (1-loop) contribution does not obey the 
Callan-Symanzik equation (\ref{Phi2CZeqn}); the $\frac{H^2}{4\pi^2} \ln(a)$
contribution is a stochastic effect which is not explained by the 
renormalization group.

Expressions (\ref{deltaA2}) and (\ref{KA2}) show that the composite operator 
$A^2(x)$ does not require a curvature-dependent field strength renormalization 
at this order,
\begin{equation}
Z_{A^2} = 1 - 2 K_{A2} \times R + O(\lambda^4) = 1 + O(\lambda^4) \quad 
\Longrightarrow \quad \gamma_{A^2} = 0 + O(\lambda^4) \; . \label{gammaA2}
\end{equation}
Hence the Callan-Symanzik equation does not constrain $\langle \Omega \vert 
A^2(x) \vert \Omega \rangle_{\rm ren}$ at this order, and we saw from
expression (\ref{stochA2VEV}) that stochastic effects completely explain the 
1-loop and 2-loop contributions for this entry in Table~\ref{DoubleLogs}.

The composite operator $B^2(x)$ experiences a curvature-dependent field strength
renormalization we can read off from expression (\ref{deltaB2}) and (\ref{KB2}),
\begin{equation}
Z_{B^2} = 1 - 2 K_{B2} \times R + O(\lambda^4) \qquad \Longrightarrow \qquad
\gamma_{B^2} = -\frac{3 \lambda^2 H^2}{16 \pi^2} + O(\lambda^4 H^4) \; .
\label{gammaB2}
\end{equation}
The Callan-Symanzik equation for $\langle \Omega \vert B^2(x) \vert \Omega
\rangle_{\rm ren}$ is,
\begin{equation}
\Bigl[ a \frac{\partial}{\partial a} + \beta \frac{\partial}{\partial \lambda}
+ \gamma_{B^2} \Bigr] \Bigl\langle \Omega \Bigl\vert B^2(x) \Bigr\vert
\Omega \Bigr\rangle_{\rm ren} = 0 \; . \label{B2CZeqn}
\end{equation}
Because the $\beta$ function of the two field model is of order $\lambda^3$,
the equation precisely predicts the $\frac{3 \lambda^2 H^2}{32 \pi^2} \times
\frac{H^2}{4 \pi^2} \ln(a)$ contribution reported in Table~\ref{DoubleLogs}.
Note again that the Callan-Symanzik equation does not predict the logarithm
at order $\lambda^0$ which is a stochastic effect. 

Expression (\ref{Cterms}) shows that we can also think of the 1PI 2-point 
functions for $A$ and $B$ experiencing curvature-dependent field strength 
renormalizations whose associated $\gamma$ functions can be computed from
(\ref{CA12}) and (\ref{CB2}),
\begin{eqnarray}
Z_A = 1 + C_{A2} \!\times\! R + O(\lambda^4) & \quad , \quad & 
Z_B = 1 + C_{B2} \!\times\! R + O(\lambda^4) \; , \label{ZAB} \\
\gamma_{A} = +\frac{\lambda^2 H^2}{32 \pi^2} + O(\lambda^4) 
& \quad , \quad & \gamma_{B} = -\frac{\lambda^2 H^2}{32 \pi^2} + 
O(\lambda^4) \; . \label{gammaAB}
\end{eqnarray}
The tree order mode functions both approach constants at late times, so the
1-loop corrections are unconstrained by the Callan-Symanzik equation. 
However, the tree order exchange potentials approach $\frac{K H}{4\pi} 
\ln(Hr)$ at late times. We must therefore interpret the $\mu 
\frac{\partial}{\partial \mu}$ term as $r \frac{\partial}{\partial r}$.
The 1-loop corrections are integrals of the 1PI 2-point functions, so
the Callan-Symanzik equations read,
\begin{equation}
\Bigl[r \frac{\partial}{\partial r} + \beta \frac{\partial}{\partial 
\lambda} - 2 \gamma_A\Bigr] P_A(\eta,r) = 0 = \Bigl[r \frac{\partial}{\partial r} 
+ \beta \frac{\partial}{\partial \lambda} - 2 \gamma_B\Bigr] P_B(\eta,r) \; .
\label{PABCZeqn}
\end{equation}
With the $\gamma$ functions (\ref{gammaAB}), these equations predict
the $\pm \frac{\lambda^2 H^2}{32 \pi^2} \ln(H r) \times \frac{K H}{4 \pi}
\ln(H r)$ contributions for $P_A(\eta,r)$ and $P_{B}(\eta,r)$ in 
Table~\ref{DoubleLogs}. Note that they do not predict the $-\frac{\lambda^2
H^2}{32 \pi^2} \ln(a) \times \frac{K H}{4 \pi} \ln(H r)$ contribution to
$P_{A}(\eta,r)$. This is a stochastic effect from the mass generated by the
effective potential $V_{\rm eff}(A)$ of expression (\ref{VA}).

\subsection{Color-Coded Tables}

So many different logarithms occurred that we have thought it good to 
provide color-coded versions of Tables~\ref{SingleLogs} and \ref{DoubleLogs} 
to distinguish stochastic effects (in red) from those explained by the 
renormalization group (in green).
\begin{table}[H]
\setlength{\tabcolsep}{8pt}
\def\arraystretch{1.5}
\centering
\begin{tabular}{|@{\hskip 1mm }c@{\hskip 1mm }||c|}
\hline
Quantity & Leading Logarithms \\
\hline\hline
$u_{\Phi}(\eta,k)$ & $\Bigl\{1 {\color{red} + \frac{\lambda^2 H^2}{32 \pi^2} \ln(a)}
+ O(\lambda^4)\Bigr\} \times \frac{H}{\sqrt{2 k^3}}$ \\
\hline
$P_{\Phi}(\eta,r)$ & $\Bigl\{1 {\color{red} + 
\frac{\lambda^2 H^2}{32 \pi^2} \ln(a) }
+ O(\lambda^4)\Bigr\} \times \frac{KH}{4\pi} \ln(Hr)$ \\
\hline
$\langle \Omega \vert \Phi(x)\vert \Omega \rangle$ & $-\Bigl\{1 
{\color{red} + \frac{15 \lambda^2 H^2}{64 \pi^2} \ln(a) } + 
O(\lambda^4)\Bigr\} \times {\color{red} \frac{\lambda H^2}{16 \pi^2} 
\ln(a)}$ \\
\hline
$\langle \Omega \vert \Phi^2(x)\vert \Omega \rangle_{\rm ren}$ & 
$\Bigl\{1 {\color{green} + \frac{15 \lambda^2 H^2}{64 \pi^2} \ln(a) }
+ O(\lambda^4)\Bigr\} \times {\color{red} \frac{H^2}{4\pi^2} \ln(a)}$ \\
\hline
\end{tabular}
\caption{\footnotesize Color-coded explanations of single scalar 
logarithms from Table~\ref{SingleLogs}. Red denotes stochastic 
logarithms and green those explained by the renormalization group.}
\label{ExplainedSingleLogs}
\end{table}
\noindent Note that mass effects are considered stochastic because 
$m^2_{\Phi}$ and $m^2_{A}$ were induced by the effective potentials 
$V_{\rm eff}(\Phi)$ and $V_{\rm eff}(A)$ which give rise to the Langevin 
equations (\ref{Langevin}) and (\ref{stochAeqn}).
\begin{table}[H]
\setlength{\tabcolsep}{8pt}
\def\arraystretch{1.5}
\centering
\begin{tabular}{|@{\hskip 1mm }c@{\hskip 1mm }||c|}
\hline
Quantity & Leading Logarithms \\
\hline\hline
$u_{A}(\eta,k)$ & $\Bigl\{1 {\color{red} -\frac{\lambda^2 H^2}{32 \pi^2} 
\ln(a) } + O(\lambda^4)\Bigr\} \times \frac{H}{\sqrt{2 k^3}}$ \\
\hline
$u_{B}(\eta,k)$ & $\Bigl\{1 + 0 + O(\lambda^4)\Bigr\} \times 
\frac{H}{\sqrt{2 k^3}}$ \\
\hline
$P_{A}(\eta,r)$ & $\Bigl\{1 {\color{red} - \frac{\lambda^2 H^2}{32 \pi^2} 
\ln(a) } {\color{green} + \frac{\lambda^2 H^2}{32 \pi^2} \ln(Hr)} + 
O(\lambda^4)\Bigr\} \times \frac{KH}{4\pi} \ln(Hr)$ \\
\hline
$P_{B}(\eta,r)$ & $\Bigl\{1 {\color{green} - \frac{\lambda^2 H^2}{32 \pi^2} 
\ln(Hr) } + O(\lambda^4)\Bigr\} \times \frac{KH}{4\pi} \ln(Hr)$ \\
\hline
$\langle \Omega \vert A(x)\vert \Omega \rangle$ & $\Bigl\{1 + 
O(\lambda^2)\Bigr\} \times {\color{red} \frac{\lambda H^2}{16 \pi^2} 
\ln(a)}$ \\
\hline
$\langle \Omega \vert A^2(x)\vert \Omega \rangle_{\rm ren}$ & 
$\Bigl\{1 {\color{red} - \frac{\lambda^2 H^2}{64 \pi^2} \ln(a) }
+ O(\lambda^4)\Bigr\} \times {\color{red} \frac{H^2}{4\pi^2} \ln(a)}$ \\
\hline
$\langle \Omega \vert B(x)\vert \Omega \rangle$ & $0$ \\
\hline
$\langle \Omega \vert B^2(x)\vert \Omega \rangle_{\rm ren}$ & 
$\Bigl\{1 {\color{green} + \frac{3 \lambda^2 H^2}{32 \pi^2} \ln(a)}
+ O(\lambda^4) \Bigr\} \times {\color{red} \frac{H^2}{4\pi^2} \ln(a)}$ \\
\hline
\end{tabular}
\caption{\footnotesize Color-coded explanations of two scalar 
logarithms from Table~\ref{DoubleLogs}. Red denotes stochastic 
logarithms and green those explained by the renormalization group.}
\label{ExplainedDoubleLogs}
\end{table}

\section{Epilogue}

Proponents of the renormalization group have long contended with supporters
of the stochastic formalism in attempting to explain and re-sum the large 
logarithms which arise when making perturbative computations during 
inflation. Although the stochastic formalism provides a complete description
for scalar potential models, the outcome for nonlinear sigma models is more
nuanced. Many of their logarithms can be explained by a variant of the 
stochastic formalism which is based on using the effective potential to 
infer a scalar potential model. Unlike the cases of Yukawa theory 
\cite{Miao:2006pn} and Scalar Quantum Electrodynamics \cite{Prokopec:2007ak}, 
the effective potentials of nonlinear sigma models derive from kinetic terms 
and would vanish in flat space background. The remaining logarithms can be 
explained by a variant of the renormalization group based on regarding 
certain higher-derivative counterterms as curvature-dependent 
renormalizations of couplings in the bare theory. 

We considered a single field model (\ref{LPhi}), which can be reduced to
a free theory by a field redefinition,\footnote{The absence of 
flat space scattering in no way precludes interactions from 
changing the kinematics of free fields or
the evolution of the background.} and a two field model (\ref{LAB})
which cannot be. For the mode functions and exchange potentials of
each model we derived tree order and 1-loop results; for the expectation
values of the fields and the squares we derived 1-loop and 2-loop
results. Tables~\ref{ExplainedSingleLogs} and \ref{ExplainedDoubleLogs} 
give a color-coded summary of which logarithms have a stochastic explanation
and which ones derive from the renormalization group. Note that many of 
the entries derive partially from one technique and partly from the
other. This is particularly evident for 1-loop corrections to the 
exchange potential for $A$ in the two field model (\ref{LAB}).

The need to combine ultraviolet and stochastic techniques can be seen
in the passage from the exact field equation (\ref{Avar1}) for $A(x)$ to
its stochastic realization (\ref{stochAeqn}). The exact Heisenberg operator
equation is,
\begin{equation}
\partial_{\mu} \Bigl( \sqrt{-g} \, g^{\mu\nu} \partial_{\nu} A\Bigr)
- \frac{\lambda}{2} \Bigl(1 \!+\! \frac{\lambda}{2} A\Bigr) \partial_{\mu}
B \partial_{\nu} B g^{\mu\nu} \sqrt{-g} = 0 \; .\label{exactAeqn}
\end{equation}
The stochastic realization of the first term in (\ref{exactAeqn}) is 
straightforward \cite{Tsamis:2005hd},
\begin{equation}
\partial_{\mu} \Bigl( \sqrt{-g} \, g^{\mu\nu} \partial_{\nu} A\Bigr)
\longrightarrow -3 H \Bigl( \dot{\mathcal{A}} - \dot{\mathcal{A}}_0
\Bigr) a^3 \; . \label{stochterm1}
\end{equation}
However, there is no completely stochastic derivation of the stochastic
realization of the second term,
\begin{equation}
- \frac{\lambda}{2} \Bigl(1 \!+\! \frac{\lambda}{2} A\Bigr) \partial_{\mu}
B \partial_{\nu} B g^{\mu\nu} \sqrt{-g} \longrightarrow + 
\frac{3 \lambda H^4}{16 \pi^2} \frac{a^3}{1 \!+\! \frac12 \lambda \mathcal{A}}
\; , \label{stochterm2}
\end{equation}
because it depends on the ultraviolet sector of $B(x)$ to produce the correct
stochastic result,
\begin{equation}
\partial_{\mu} B(x) \partial_{\nu} B(x) \longrightarrow -\frac{(\frac{D-1}{D}) 
k H^2 g_{\mu\nu}(x)}{[1 \!+\! \frac12 \lambda \mathcal{A}(x)]^2} \; .
\label{problematic}
\end{equation}
Any stochastic truncation of $B(x)$ would result in ultraviolet finite fields
whose expectation values could never reproduce the indefinite signature so
evident in expression (\ref{problematic}). Note also that it can be {\it the 
same field} whose ultraviolet must be integrated out in derivative terms to 
give the appropriate Langevin equation. This is evident for the single field 
model (\ref{LPhi}) in the passage from the exact field equation (\ref{varPhi1}),
\begin{equation}
\Bigl(1 \!+\! \frac{\lambda}{2} \Phi\Bigr)^2 \partial_{\mu} \Bigl( \sqrt{-g} \,
g^{\mu\nu} \partial_{\nu} \Phi\Bigr) + \frac{\lambda}{2} \Bigl(1 \!+\! 
\frac{\lambda}{2} \Phi\Bigr) \partial_{\mu} \Phi \partial_{\nu} \Phi g^{\mu\nu}
\sqrt{-g} = 0 \; , \label{exactPhieqn}
\end{equation}
to its stochastic realization (\ref{Langevin}),
\begin{equation}
-\Bigl(1 \!+\! \frac{\lambda}{2} \varphi\Bigr)^2 \!\times\! 3 H a^3 \Bigl(
\dot{\varphi} - \dot{\varphi}_0\Bigr) - \frac{3 \lambda H^4}{16 \pi^2} 
\frac{a^3}{ 1 \!+\! \frac12 \lambda \varphi} = 0 \; . \label{stochPhieqn}
\end{equation}

Quantum field theories in an expanding universe have instantaneous
energy eigenstates, but the expansion of the universe prevents these 
eigenstates from evolving onto one another. So what was the minimum energy
state at one instant is not generally minimum energy later on. Bunch-Davies
vacuum corresponds to the state that was minimum energy in the distant past.
Although Starobinsky's stochastic formalism, and hence also our variant of it, 
was derived assuming quantum fields in Bunch-Davies vacuum, it ought to apply 
broadly to states which are perturbatively nearby. On the other hand, this is 
not true for states which are highly excited from Bunch-Davies vacuum. Indeed, 
by making suitable Bogoliubov transformations one can change the scalar 
and tensor power spectra by potentially momentum-dependent factors which range 
from zero to infinity! That same ambiguity must also afflict the stochastic 
formalism, as it does all the other predictions of inflationary cosmology. We
suspect that the process by which the states of originally trans-Planckian wave
numbers are red-shifted to the point where quantum general relativity can be
used as an effective field theory leaves these states near Bunch-Davies vacuum.
However, that is a conjecture which can and should be studied.

We should comment on the peculiar notion of applying the renormalization 
group to nonrenormalizable theories such as (\ref{LPhi}) and (\ref{LAB}). 
At the order we have worked, no renormalization was required for the 1PI 
2-point functions of the single field model (\ref{LPhi}), but we did need 
four counterterms (\ref{Cterms}) to renormalize the self-masses for $A$ 
and $B$,
\begin{eqnarray}
\lefteqn{\Delta \mathcal{L} = -\frac12 C_{A1} \square A \square A \sqrt{-g} 
- \frac12 C_{A2} \, R \partial_{\mu} A \partial_{\nu} A g^{\mu\nu} \sqrt{-g} 
} \nonumber \\
& & \hspace{3.4cm} - \frac12 C_{B1} \square B \square B \sqrt{-g} - \frac12 
C_{B2} \, R \partial_{\mu} B \partial_{\nu} B g^{\mu\nu} \sqrt{-g} \; . 
\qquad
\end{eqnarray}
Each of these counterterms involves higher derivatives, but there is an
important distinction between when those derivatives act on $A$ and $B$
and when they act on the metric. The terms proportional to $C_{A1}$ and 
$C_{B1}$ involve higher derivatives of the fields $A$ and $B$ and play no 
role in the generation of large inflationary logarithms. However, the terms 
proportional to $C_{A2}$ and $C_{B2}$ can be viewed as curvature-dependent 
field strength renormalizations of $A$ and $B$, respectively. It is the 
flow of these couplings which serves to capture the green-colored logarithms 
in Tables~\ref{ExplainedSingleLogs} and \ref{ExplainedDoubleLogs}. It is
also worth noting that the basis for stochastic effects, the effective 
potentials (\ref{VPhi}) and (\ref{VA}), are also curvature-dependent and 
would vanish in the flat space limit.

This project suggests a number of extensions. The most urgent of these is
working out the curvature-dependent coupling constant renormalizations that
would allow us to determine the renormalization group flows. We would like
to determine the late time behavior and also whether or not renormalization
group improvement of the effective potentials matters at leading logarithm 
order. It would also be interesting to compute the order $\lambda^3$ (two 
loop) contribution to the expectation value of $A(x)$ to see if it agrees
with the stochastic prediction in expression (\ref{stochAVEV}). And we 
would like to know whether or not the renormalization group can be used to 
explain the sub-dominant logarithms one sometimes encounters in the rate at 
which the mode functions freeze in,
\begin{eqnarray}
u_0(\eta,k) & \longrightarrow & \frac{H}{\sqrt{2 k^3}} \Bigl\{1 + 
\frac{k^2}{2 a^2 H^2} + \dots \Bigr\} \; , \\
u_1(\eta,k) & \longrightarrow & \frac{H}{\sqrt{2 k^3}} \Bigl\{0 + 
\frac{\# \lambda^2 k^2 \ln(a)}{a^2} + \dots \Bigr\} \; .
\end{eqnarray}
A final spin-off is understanding the painfully accumulated collection of 
large logarithms induced by inflationary gravitons in the mode functions 
and exchange potentials of various matter theories \cite{Miao:2006gj,
Glavan:2013jca,Wang:2014tza,Glavan:2021adm} and gravity itself 
\cite{Park:2015kua,Tan:2021lza}.

Of course the primary motivation for studying nonlinear sigma models was
to understand the derivative interactions of quantum gravity, without the
plethora of indices and the miasma of confusion associated with gauge 
fixing. We particularly wish to understand the viability of back-reaction
to slow the expansion rate in $\Lambda$-driven inflation \cite{Tsamis:1996qq,
Tsamis:2011ep}. It is therefore worth summarizing what this project suggests 
for quantum gravity:
\begin{itemize}
\item{The large logarithms induced by inflationary gravitons can likely be 
explained using a combination of curvature-dependent effective potentials 
and curvature-dependent renormalization group flows;}
\item{There seems to be no obstacle to inferring an effective potential by
integrating the ultraviolet out of the invariant Lagrangian 
\cite{Tsamis:1992xa},
\begin{eqnarray}
\lefteqn{ \mathcal{L}_{\rm inv} = a^{D-2} \sqrt{-\widetilde{g}} \, 
\widetilde{g}^{\alpha\beta} \widetilde{g}^{\gamma\delta} \widetilde{g}^{
\epsilon\zeta} \Bigl[ \frac12 h_{\alpha \gamma , \epsilon} h_{\zeta \delta , 
\beta} \!-\! \frac12 h_{\alpha \beta , \gamma} h_{\delta \epsilon , \zeta} 
\!+\! \frac14 h_{\alpha \beta , \gamma} h_{\epsilon \zeta ,\delta} }
\nonumber \\ 
& & \hspace{1cm} - \frac14 h_{\alpha \gamma , \epsilon} h_{\beta \delta , 
\zeta} \Bigr] + \frac12 (D\!-\!2) a^{D-1} H \sqrt{-\widetilde{g}} \,
\widetilde{g}^{\alpha\beta} \widetilde{g}^{\gamma\delta} h_{\alpha\beta , 
\gamma} h_{\delta 0} \; , \qquad \label{Linv}
\end{eqnarray}
where $\widetilde{g}_{\mu\nu} \equiv \eta_{\mu\nu} + \kappa h_{\mu\nu}$ is
considered to be constant ($\kappa^2 \equiv 16 \pi G$), at which point we
can follow back-reaction by solving for the homogeneous evolution with the
effective potential, the same way we did with equations 
(\ref{Phievolution}-\ref{classPhi}) for the single field model and with 
equations (\ref{Aevolution}-\ref{Asoln}) for the two field model;}
\item{Of the two invariant 1-loop counterterms,
\begin{equation}
\Delta \mathcal{L} = \alpha_1 R^2 \sqrt{-g} + \alpha_2 C^{\alpha\beta\gamma
\delta} C_{\alpha\beta\gamma\delta} \sqrt{-g} \; ,
\end{equation}
the one proportional to $\alpha_2$ likely plays no role in producing large
logarithms while the one proportional to $\alpha_1$ can be viewed as a 
curvature-dependent renormalization of Newton's constant, and its flow has
the potential to explain the unnaturally large value of $\alpha_1$ in
Starobinsky's original model of inflation \cite{Starobinsky:1980te};}
\item{As long as the curvature remains nonzero there is no reason to assume 
that evolution approaches a static limit, and the two field model provides
an explicit example of significant evolution persisting to arbitrarily late
times, cf. expression (\ref{Asoln});}
\item{It seems inevitable that significant modifications to the expansion 
rate and to the force of gravity will persist to late times; and}
\item{Curvature-dependent effective potentials and renormalization group 
flows pose a challenge when back-reaction causes the curvature to change, but 
they also provide a natural mechanism through which the effects of inflationary
gravitons can become dormant during radiation domination (with $R = 0$) and 
then reassert themselves at late times, after the transition to matter 
domination.}
\end{itemize}

\vspace{.5cm}

\centerline{\bf Acknowledgements}

This work was partially supported by Taiwan MOST grants 109-2112-M-006-002
and 110-2112-M-006-026; by the European Union's Horizon 2020 Programme under 
grant agreement 669288-SM-GRAV-ERC-2014-ADG; by NSF grant 1912484; and by the 
Institute for Fundamental Theory at the University of Florida.

\end{document}